\documentclass[a4paper,twocolumn,showpacs,superscriptaddress,floatfix,nofootinbib,prd]{revtex4-1}
\usepackage{graphicx}
\usepackage{epsf}
\usepackage{epsfig}
\usepackage{amssymb,amsmath}
\usepackage[usenames]{color}
\usepackage{amssymb}
\usepackage{wasysym}
\usepackage{times}
\usepackage{mathrsfs}
\usepackage{hyperref}
\usepackage[normalem]{ulem}
\hypersetup{
  colorlinks=true,        
  linkcolor=blue,         
  citecolor=cyan,         
}
\usepackage{float}
\definecolor{orange}{rgb}{1,0.5,0}

\newcommand{\cf}{cf.~}
\newcommand{\ie}{i.e.,~}
\newcommand{\eg}{e.g.,~}
\newcommand{\ttt}[1]{\texttt{\small #1}}
\renewcommand{\BibitemShut}[1]{}
\newcommand\pasa{\ref@jnl{PASA}}

\newcommand{\km}{\,{\mathrm{km}}}

\newcommand{\Msun}{M_{\odot}}
\newcommand{\Msol}{M_{\odot}}
\newcommand{\Ye}{Y_\mathrm{e}}
\newcommand{\kB}{k_\mathrm{B}}
\graphicspath{{figs/}}
\begin{document}

\title[Eccentric binaries]{Dynamical ejecta and nucleosynthetic yields
  from eccentric binary neutron-star mergers}

\author{L.~Jens~Papenfort}\thanks{E-mail: papenfort@th.physik.uni-frankfurt.de}
\affiliation{Institut f{\"u}r Theoretische Physik,
Johann Wolfgang Goethe-Universit\"at, Max-von-Laue-Stra{\ss}e 1, 60438 Frankfurt,
Germany}
\author{Roman~Gold}
\affiliation{Institut f{\"u}r Theoretische Physik,
Johann Wolfgang Goethe-Universit\"at, Max-von-Laue-Stra{\ss}e 1, 60438 Frankfurt,
Germany}
\author{Luciano~Rezzolla}
\affiliation{Institut f{\"u}r Theoretische Physik,
Johann Wolfgang Goethe-Universit\"at, Max-von-Laue-Stra{\ss}e 1, 60438 Frankfurt,
Germany}
\affiliation{Frankfurt Institute for Advanced Studies,
Ruth-Moufang-Stra{\ss}e 1, 60438 Frankfurt, Germany}

\begin{abstract}
With the recent advent of multi-messenger gravitational-wave astronomy
and in anticipation of more sensitive, next-generation gravitational-wave
detectors, we investigate the dynamics, gravitational-wave emission, and
nucleosynthetic yields of numerous eccentric binary neutron-star mergers
having different equations of state. For each equation of state we vary
the orbital properties around the threshold of immediate merger, as well
as the binary mass ratio. In addition to a study of the
gravitational-wave emission including $f$-mode oscillations before and
after merger, we couple the dynamical ejecta output from the simulations
to the nuclear-reaction network code \texttt{SkyNet} to compute
nucleosynthetic yields and compare to the corresponding results in the
case of a quasi-circular merger. We find that the amount and velocity of
dynamically ejected material is always much larger than in the
quasi-circular case, reaching maximal values of $M_{\rm ej, max} \sim 0.1
\, \Msol$ and $v_{\rm max}/c \sim 0.75$. At the same time, the properties
of this material are rather insensitive to the details of the orbit, such
as pericenter distance or post-encounter apoastron distance.
Furthermore, while the composition of the ejected matter depends on the
orbital parameters and on the equation of state, the relative
nucleosynthetic yields do not, thus indicating that kilonova signatures
could provide information on the orbital properties of dynamically
captured neutron-star binaries.
\end{abstract}


\maketitle

\section{Introduction}
\label{sec:intro}

The emission of gravitational waves (GWs) emitted from binary
neutron-star systems during inspiral as well as the electromagnetic
signals emitted after merger have entered the observational realm with
the detection of GW170817 \cite{Abbott2017,Abbott2017b}. This event not
only signalled the beginning of multimessenger astronomy exploiting
gravitational-wave observations but also confirmed the hypothesised link
between compact binary mergers and short gamma-ray bursts
\cite{Narayan92, Eichler89, Rezzolla:2011, Paschalidis2016,
  Metzger2017}. Numerical simulations of merging neutron stars (see
\cite{Baiotti2016, Paschalidis2016} for recent reviews), which now
include also the modelling of the kilonova emission from this process
\cite{Bovard2017, Metzger2017, Perego2017}, together with multimessenger
observations, have provided important insight on the maximum mass of
neutron stars and on the expected distribution in radii \cite{Annala2017,
  Bauswein2017b, Margalit2017, Radice2017b, Rezzolla2017, Ruiz2017,
  Shibata2017c, Most2018}.

Most work on binary neutron stars has focused on the primordial
population where the binaries decouple from their environments at large
separation and evolve as a two-body system in isolation. In such a
configuration, binary orbits efficiently circularise due to GW emission
\cite{Peters:1964} and therefore eccentricity can be eliminated as a free
parameter of the system. The binary sources detected by advanced LIGO
(adLIGO) are consistent with a low but finite binary eccentricity of
${\rm e} \lesssim 0.1$ when entering the adLIGO frequency range
\cite{Huerta2016, Huerta2017, Yang2018}.

The independent population of dynamically formed binaries bears little
resemblance to the primordial population. While in both cases the neutron
stars are born in supernova explosions, their orbital characteristics are
very different. In astrophysical environments where the stellar density
is high, such as globular clusters or galactic cores, the assumptions
made in Refs. \cite{Peters:1963ux,Peters:1964} do not apply. Instead,
$N$-body effects have to be taken into account to understand typical
parameters and event rates (see \cite{Samsing2016, Rodriguez2017,
  Bonetti2018} for recent work). Calculations of event rates for binary
mergers are challenging even in the primordial case, mainly because
many different physical conditions affect the system throughout its long
life and many of the involved processes, such as the common-envelope
phase, are complicated; despite impressive efforts \cite{Ivanova2013},
the combination of these effects remain poorly understood. In addition to
these challenges, event rates in dynamically formed systems also depend
on the size of the star cluster, on the binary fraction, as well as on
the hardening and evaporation, just to name a few additional
complications. As a result, event-rate estimates for dynamically-formed
binaries are even more uncertain. Bearing these considerations in mind,
the estimated event rates are roughly in the range of $0.003-6\,{\rm
  Gpc}^{-3}\,{\rm yr}^{-1}$ \cite{Tsang2013}. It is therefore important
to investigate the range out to which such sources can be seen. We focus
here on the intrinsic aspects related to the dynamics of the system and
GW emission. For the sensitivity and instrumental characteristics of
adLIGO the numbers suggest so far that GW detections from eccentric
sources are expected to be rare at best. In particular, highly eccentric
binary neutron stars are extremely difficult to detect for the detector
characteristics of adLIGO \cite{Tan2017} mainly due to missing signal
power at frequencies around $\sim 100\,{\rm Hz}$ where the detector
performs best.

The situation is different for next-generation GW detectors such as the
Einstein Telescope (ET) \cite{Punturo:2010, Punturo2010b} or the Cosmic
Explorer (CE) \cite{Evans2016}. In this new detector era, substantially
higher sensitivities will enable detections from cosmological
distances. Apart from the naive gain in observed volume, which already
boosts detection rates considerably, signals from redshifts $z=3$ (as
applicable to ET) or even $z=6$ (as applicable to CE) are shifted
substantially towards lower frequencies by a factor $1/(1+z)$. The
technological advances that will accompany these instruments, together
with novel concepts to use unstable optomechanical filters to compensate
for phase dispersion \cite{Miao2015, Miao2017} and thus increase the
sensitivities of current interferometers near $2\,{\rm kHz}$ to levels
comparable to ET, make high-frequency GW signals such as merger
waveforms, ringdown and $f$-mode oscillations much more detectable. For a
combined ET and CE era, $0.1-10$ eccentric binary neutron-star captures
could be detected per year \cite{Chirenti:2016b}.

The detection of and parameter estimation from GWs emitted by eccentric
binaries is not only an instrumental issue, but also a challenge from a
data analysis and modelling perspective. Past and current searches
performed on adLIGO data (see, \eg \cite{Bose2017}) are ill-suited for
the sources discussed here and one promising suggestion is to use
power-stacking \cite{Tai2014} to boost the SNR. If eccentric sources are
modeled accurately enough, then eccentric binaries can be seen out to
larger distances, and parameters such as chirp mass and sky localisation
can be estimated more accurately \cite{Gondan2017}, mostly because the
richer structure in eccentric waveforms effectively breaks parameter
degeneracies. Therefore, even just one detection of GWs from eccentric
binaries offers more scientific opportunities than many quasi-circular
binary signals.

The richer structure of GWs from eccentric binaries ranges from
small-amplitude modulations in the GW signal when the binary has also
modest eccentricities, to short-term GW bursts emitted during each
pericenter passage of the binary, which are then separated by a phase
with no significant gravitational radiation, in the case of binaries with
large eccentricities.  It was first realised in Refs. \cite{Turner1977a,
  Turner1977b}, that in case of binaries containing at least one neutron
star, there is an additional effect between the pericenter bursts: Tidal
perturbations during pericenter passage can excite stellar $f$-mode
oscillations with a time-varying quadrupole moment and that themselves
act as sources of gravitational radiation (see also \cite{Parisi2017} and
\cite{Chaurasia2018} for a very recent analysis). This $f$-mode signal
can dominate the GW losses away from pericenter. As a consequence, the
instantaneous GW frequency transitions from the binary-dominated burst
signal over to the star-dominated, constant $f$-mode frequency. These
$f$-mode oscillations drain energy from the orbit, causing a GW phase
shift and depend on the internal structure and equation of state (EOS) of
the neutron star in a way which is not too different from the post-merger
signal (see, \eg \cite{Bauswein2011, Stergioulas2011b, Takami2014,
  Takami2015, Bernuzzi2015a, Rezzolla2016}).

Initial nonlinear studies employing global hydrodynamical simulations in
time-dependent spacetimes of dynamically-formed neutron-star binaries
have not only confirmed earlier predictions for and quantified the
$f$-mode contributions, but have further shown that eccentricity also has
substantial consequences for the matter dynamics in the late inspiral,
merger and post-merger phase of the system \cite{Stephens2011, Gold2012,
  Moldenhauer2014, Paschalidis2015, East2016, East2016a}, giving rise to
a much richer phenomenology compared to the quasi-circular case (see also
\cite{East2012b0} for a study on black hole-neutron-star systems). In
studies that feature the formation of a hypermassive neutron star (HMNS)
it was discovered that generically (for all EOSs considered) a one-armed
spiral instability occurs, that produces gravitational radiation mainly
in the $\ell=2,m=1$ mode \cite{Paschalidis2015, East2016, East2016a,
  Lehner2016a}, although this contribution is much smaller than the one
in the traditional $\ell=2,m=2$ mode \cite{Radice2016a, Hanauske2016}.

Aside from GW emission, the dynamical ejecta\footnote{We refer to as
  \textit{``dynamical''} ejecta the unbound part of the neutron-star
  matter that is lost because of tidal interactions and shock heating
  over a timescale of the order of tens of milliseconds, hence comparable
  with the dynamical timescale of the system. This is to be constrasted
  with the \textit{``secular''} ejecta, which is instead lost on a
  timescale of about one second via neutrino or magnetically driven
  winds.} of neutron-rich matter and hence nucleosynthetic yields can be
expected to be greatly affected by orbital eccentricity
\cite{Radice2016}. It is therefore important to investigate to what
extent nucleosynthesis depends on orbital eccentricity in the context of
fast mergers in the early universe \cite{Ji2016,Bonetti2018}. Such
differences would lead to different implications for the
near-infrared/optical lightcurves emitted from such systems
\cite{Tanvir2013, Wollaeger2017, Bovard2017}.

In this paper we report on the most extensive to date investigation of
eccentric binary neutron-star evolutions in full general relativity. We
focus on following the ejected mass with tracers, compute the resulting
nucleosynthetic yields, analyze the spectrograms of the GWs, and compare
our results obtained with two different nuclear EOSs and different mass
ratios.  We show that the mass outflow from these systems is
systematically larger than in the corresponding case of binaries with the
same EOS and total mass, but in quasi-circular orbits, reaching extreme
values of almost $0.1 \, \Msol$ of dynamically ejected material. In fact,
even the ejected matter from a single non-merging encounter can easily
exceed the ejected mass of an entire quasi-circular inspiral and
merger. In addition, the more massive ejecta reach higher maximal
velocities and both aspects can have important implications for
generating kilonova lightcurves \cite{Grossman2014}. Despite these
striking differences in the ejecta properties when compared to their
quasi-circular counterparts, we find that the relative nucleosynthetic
yields from dynamical ejecta are rather insensitive to the orbital
eccentricity\footnote{Hereafter all of our considerations will refer to
  the ``dynamical'' ejecta and we note that such considerations may be
  subject to modifications when a complete picture, including ``secular''
  ejecta, can be handled through long-term numerical simulations that are
  still computationally prohibitive.}.

We also consider the GW emission from these systems and show that the
overall structure of pre-merger GW signals consisting of $f$-mode
oscillations, that are triggered at the periastron passages accompanied
by strong bursts, are rather insensitive to the equation of state. Our
findings indicate that the systems under investigation here are
detectable out to redshifts as large as $z\sim 0.5$ for ET and CE, but
other plausible, not too dissimilar configurations could be detected out
to larger distances. In general, our study points to the fact that longer
signals with more cycles and/or multiple bursts are more easily detected
than short signals with only a single encounter. Although these mergers
are not expected to be very common, we argue that, peculiar factors like
low-frequency contributions from $f$-modes, cosmological redshift of GW
signals, better parameter estimation, and the resulting increased
detector volume in the context of future GW detectors have to be factored
in when estimating their event rates. A final judgement of the true event
rates of such systems therefore awaits further detailed and quantitative
study of these aspects.

The plan of the paper is as follows. In Sec. \ref{sec:methods} we recall
the mathematical and numerical set-up employed in our simulations,
together with our choice of EOS and initial data. In Sec.
\ref{sec:results} we present instead our findings, concentrating on the
orbital dynamics (\ref{sub:orbit-dynamics}), on the mass ejection
(\ref{sec:me}), on the properties of the dynamical ejecta,
(\ref{sub:dynejec-props}), on the r-process nucleosynthesis
(\ref{sec:nucleo}), and on the GW emission (\ref{sec:ge}). Finally,
Sec. \ref{sec:conclusions} contains our conclusions and the prospects of
future work.

\section{Methods}
\label{sec:methods}

In this section we describe the most important parts composing our
numerical setup. This includes the evolution schemes, the initial
data and data analysis.

\subsection{Spacetime evolution}

The spacetime evolution is governed by the Einstein field equations
\begin{equation}
  R_{\mu\nu}-\frac{1}{2}R g_{\mu\nu} = 8\pi T_{\mu\nu}\,,
  \label{eq:EFE}
\end{equation}
for the four-dimensional metric $g_{\mu\nu}$. As usual these
equations are cast into a form suitable for a numerical treatment as a
Cauchy Initial-Value Problem adopting the standard 3+1 split of
spacetime \cite{Alcubierre:2008, Rezzolla_book:2013}. In this
formalism, the metric takes the form

\begin{equation}
  ds^{2} = -\alpha^{2} dt^{2} + \gamma_{ij} (dx^{i} + \beta^{i} dt)(dx^{j} + \beta^{j} dt)\,,
  \label{eq:3p1_metric}
\end{equation}
with $\gamma_{ij}$ being the spatial metric induced on each hypersurface
of the spacetime foliation, $\alpha$ the lapse function and $\beta^{i}$
the shift vector. Together, they form the spatial hypersurface unit
normal vector
\begin{equation}
  n^{\mu}=\frac{1}{\alpha}\left(1,-{\beta^{i}}\right)  \,.
\end{equation}
Using this representation of a general spacetime, it is possible to
decompose the Einstein field equations into spatial and temporal
parts. Different choices for the independent variables and the addition
of multiples of the constraint equations leads to different formulations
that are equivalent in the continuum but have different properties in a
numerical simulation with finite temporal and spatial resolution.  Here,
we employ the CCZ4 formalism \cite{Alic:2011a, Alic2013, Bezares2017}
(see also \cite{Dumbser2017} for a first-order version), for which the
starting point is the covariant extension of the Einstein equations
\begin{eqnarray}
&& \hskip -0.5cm 8\pi\left(T_{\mu\nu}  - \frac{1}{2}g_{\mu\nu}T\right) = R_{\mu\nu} +
2\nabla_{(\mu}Z_{\nu)} \nonumber \\
&& \hskip 1.5cm
+ \kappa_{1}\left[2n_{(\mu}Z_{\nu)}-\left(1+\kappa_{2}\right)g_{\mu\nu}n_{\sigma}Z^{\sigma}\right]
\,,
\label{eq:CCZ4}
\end{eqnarray}
also known as the Z4 formulation
\cite{Bona:2003fj,Bona:2003qn,Alic:2009}, where the four-vector $Z_{\mu}$
consists of the four constraint equations given by projections of
Eq.~\eqref{eq:EFE}. These constraints have to be fulfilled by any
solution of Eqs.~\eqref{eq:EFE} and in particular by the initial data for
the time integration, so if
\begin{equation}
  Z_{\mu} = 0\,,
\label{eq:Z0}
\end{equation}
the Einstein field equations are recovered. This equivalence in the
continuum limit implies that solutions to Eq.~\eqref{eq:CCZ4} under the
constraint \eqref{eq:Z0} are also solutions to Eqs.~\eqref{eq:EFE}, but
the two formulations have distinct principal parts and hence different
mathematical properties on the continuum level, some of which carry over
to a numerical discretization of Eq.~\eqref{eq:CCZ4}. The additional
terms involving the constraint-damping coefficients $\kappa_{1}$ and
$\kappa_{2}$ lead to an exponential damping of the constraint violations
on a characteristic timescale inversely proportional to the coefficients,
which is similar to the divergence cleaning approach in
magnetohydrodynamics \cite{Dedner:2002}. While constraint-violation
damping is a distinguishing feature of the CCZ4 formulation, it is
especially valuable for the approximate initial data used in this study,
whose initial violations are large and need to be reduced rapidly (see
Sec.~\ref{sec:initial-data}). In addition, even in the zero damping case,
\ie $\kappa_1=\kappa_2=0$, Z4-based formulations feature a
\emph{propagating} mode associated with the Hamiltonian constraint
violations as opposed to the classic BSSNOK formulation \cite{Shibata95,
  Baumgarte99}, where a ``zero-speed'' mode leads to local build-up of
Hamiltonian constraint violations due to numerical error. In Z4-based
formulations such violations propagate and therefore have a chance to
leave the grid. It is therefore expected that Z4-based formulations,
which have intrinsically smaller violations of the constraints, yield
more accurate spacetime evolutions than the BSSNOK system.

The evolution of the gauge conditions, \ie lapse and shift, is governed
by standard 1+log slicing condition and a Gamma driver (see \eg
\cite{Alcubierre:2008, Baumgarte2010, Rezzolla_book:2013}). For the
numerical solution of the Einstein equations we make use of the publicly
available \texttt{McLachlan} code \cite{Brown2007b} which implements the
above evolution equations with fourth-order central finite-difference
methods and is part of the \texttt{Einstein Toolkit}
\cite{Loffler:2011ay}. The nonlinear stability of the spacetime evolution
is ensured by adding an artificial Kreiss-Oliger dissipation
\cite{Kreiss73} to the spacetime variables.

\subsection{General-relativistic hydrodynamics}
\label{sec:grhydro}

The neutron stars are modeled via a relativistic perfect-fluid
energy-momentum tensor \cite{Rezzolla_book:2013}
\begin{equation}
  T_{\mu\nu} = \rho h u_{\mu} u_{\nu} + p g_{\mu\nu}\,,
\end{equation}
where $u^{\mu}$ is the four-velocity, $\rho$ the rest-mass density, $h$
the specific enthalpy, and $p$ the pressure of the fluid. The
relativistic hydrodynamic equations can be derived by means of the
Bianchi identity of the Einstein equations~\eqref{eq:EFE} giving the
local conservation law for the energy-momentum tensor
\begin{equation}
  \nabla_{\mu}T^{\mu\nu} = 0\,,
  \label{eq:em-conservation}
\end{equation}
while the conservation of rest mass is imposed via the continuity equation
\begin{equation}
  \nabla_{\mu}\left(\rho u^{\mu}\right) = 0\,.
  \label{eq:rm-conservation}
\end{equation}
The relativistic-hydrodynamic equations \eqref{eq:em-conservation} and
\eqref{eq:rm-conservation} are closed by a suitable EOS expressing the
pressure in terms of other thermodynamical variables of the fluid and
that we will discuss in more detail in Sec. \ref{sec:EOS}.

These equations can be cast into conservative form, which is also
known as the ``Valencia formulation'' \cite{Banyuls97}
\begin{equation}
  \partial_t \boldsymbol{U} + \partial_i \boldsymbol{F}^{i} = \boldsymbol{S}\,,
\end{equation}
with the conserved variables $\boldsymbol{U}$, the fluxes
$\boldsymbol{F}^{i}$ and the sources $\boldsymbol{S}$ depend on the
primitive hydrodynamic variables as well as the geometry of the spatial
hypersurfaces (see, \eg \cite{Rezzolla_book:2013} for the explicit form
of $\boldsymbol{U}$, $\boldsymbol{F}^{i}$ and $\boldsymbol{S}$).

These conservation equations are implemented in the \texttt{WhiskyTHC}
code \cite{Radice2013b, Radice2013c, Radice2015}, which uses either
finite-volume or high-order finite-differencing high-resolution
shock-capturing methods.

In these simulations we make use of a particular combination of the
finite-volume scheme, the high-order MP5 primitive reconstruction
\cite{suresh_1997_amp, Radice2012a}, the second-order HLLE Riemann solver
\cite{Harten83}, the positivity preserving limiter of Refs.
\cite{Hu2013,Radice2013c}, and, additionally, refluxing at the refinement
boundaries to further improve mass conservation \cite{Berger89}.
Finally, as standard in this type of approaches
\cite{Rezzolla_book:2013}, in order to accurately track regions that
transition from the fluid to the vacuum regime, all simulations have an
artificial low-density background atmosphere which is evolved as
discussed in \cite{Radice2013c}. For all simulations presented in this
study we choose an atmosphere value of $\rho_{\rm atmo} \sim 6 \times
10^3 \, {\rm g\,cm}^{-3}$, which is therefore of about 11 orders of
magnitude smaller than the maximum rest-mass density in the simulations.

\subsection{Neutrino losses}

Neutrino emission is modeled by a grey (\ie frequency-independent)
neutrino leakage scheme (see, \eg \cite{vanRiper1981, Ruffert97,
  Rosswog:2003, OConnor10, Perego2014}). In this scheme the trapped
neutrinos do not individually contribute to the energy-momentum tensor
itself, however the radiative losses by the non-trapped neutrinos are
modeled by a source term to the local conservation law
\eqref{eq:em-conservation} and lepton number $n_{e}$ conservation leading
to modified conservation equations defined by
\begin{align}
  \nabla_{\mu}T^{\mu\nu} &= \Psi^{\nu}\,,\\
  \nabla_{\mu}\left(n_{e} u^{\mu}\right) &= R \,,
  \label{eq:ln-conservation}
\end{align}
together with Eq.~\eqref{eq:rm-conservation}. Assuming that the neutrinos
are in the optically thick regime and thus in local thermodynamic
equilibrium with the baryonic matter, the energy-averaged approximations
to the source terms $\Psi^{\nu}$ and $R$ are computed. This regime is
matched to the optically thin regime outside the neutrinosphere by a
free-streaming approximation.  The details of this procedure have been
described in \cite{Galeazzi2013,Radice2016}. We recall that the leakege
scheme can only account for radiative losses, i.e., cooling of the dense
matter, and it does not incorporate potential neutrino absorption in the
ejecta, which would need to be modeled by a more sophisticated approach
like a M1-scheme \cite{Foucart2015, Sekiguchi2015}. On the other hand,
this approach is well justified to model the scenario under investigation
since the dynamical ejecta is cold and not strongly affected by neutrino
absorption.

\begin{table*}[t]
\begin{tabular}{c|c|c|c|c|c|c|c|c|c|c|c}
Model & EOS & $r_{p}\,\left[M_{\odot}\right]$ & $q$ & $M_{1}\,\left[M_{\odot}\right]$ & $M_{2}\,\left[M_{\odot}\right]$ & $R_{1}\,\left[\text{km}\right]$ & $R_{2}\,\left[\text{km}\right]$ & $M_{\text{b},1}\,\left[M_{\odot}\right]$ & $M_{\text{b},2}\,\left[M_{\odot}\right]$ & ${\cal C}_{1}$ & ${\cal C}_{2}$\tabularnewline
\hline
\hline
\ttt{LS220-RP11.250} & LS220 & $11.250$ & $1.0$ & $1.35$ & $1.35$ & $12.78$ & $12.78$ & $1.47$ & $1.47$ & $0.156$ & $0.156$\tabularnewline
\hline
\ttt{LS220-RP12.500} & LS220 & $12.500$ & $1.0$ & $1.35$ & $1.35$ & $12.78$ & $12.78$ & $1.47$ & $1.47$ & $0.156$ & $0.156$\tabularnewline
\hline
\ttt{LS220-RP13.000} & LS220 & $13.000$ & $1.0$ & $1.35$ & $1.35$ & $12.78$ & $12.78$ & $1.47$ & $1.47$ & $0.156$ & $0.156$\tabularnewline
\hline
\ttt{LS220-RP13.250} & LS220 & $13.250$ & $1.0$ & $1.35$ & $1.35$ & $12.78$ & $12.78$ & $1.47$ & $1.47$ & $0.156$ & $0.156$\tabularnewline
\hline
\ttt{LS220-RP13.340} & LS220 & $13.340$ & $1.0$ & $1.35$ & $1.35$ & $12.78$ & $12.78$ & $1.47$ & $1.47$ & $0.156$ & $0.156$\tabularnewline
\hline
\ttt{LS220-RP13.375} & LS220 & $13.375$ & $1.0$ & $1.35$ & $1.35$ & $12.78$ & $12.78$ & $1.47$ & $1.47$ & $0.156$ & $0.156$\tabularnewline
\hline
\hline
\ttt{LS220-RP11.250-q09} & LS220 & $11.250$ & $0.9$ & $1.35$ & $1.22$ & $12.78$ & $12.85$ & $1.47$ & $1.31$ & $0.156$ & $0.14$\tabularnewline
\hline
\ttt{LS220-RP12.500-q09} & LS220 & $12.500$ & $0.9$ & $1.35$ & $1.22$ & $12.78$ & $12.85$ & $1.47$ & $1.31$ & $0.156$ & $0.14$\tabularnewline
\hline
\ttt{LS220-RP13.000-q09} & LS220 & $13.000$ & $0.9$ & $1.35$ & $1.22$ & $12.78$ & $12.85$ & $1.47$ & $1.31$ & $0.156$ & $0.14$\tabularnewline
\hline
\hline
\ttt{DD2-RP11.250} & DD2 & $11.250$ & $1.0$ & $1.35$ & $1.35$ & $13.23$ & $13.23$ & $1.47$ & $1.47$ & $0.151$ & $0.151$\tabularnewline
\hline
\ttt{DD2-RP12.500} & DD2 & $12.500$ & $1.0$ & $1.35$ & $1.35$ & $13.23$ & $13.23$ & $1.47$ & $1.47$ & $0.151$ & $0.151$\tabularnewline
\hline
\ttt{DD2-RP13.000} & DD2 & $13.000$ & $1.0$ & $1.35$ & $1.35$ & $13.23$ & $13.23$ & $1.47$ & $1.47$ & $0.151$ & $0.151$\tabularnewline
\hline
\ttt{DD2-RP13.250} & DD2 & $13.250$ & $1.0$ & $1.35$ & $1.35$ & $13.23$ & $13.23$ & $1.47$ & $1.47$ & $0.151$ & $0.151$\tabularnewline
\hline
\ttt{DD2-RP13.375} & DD2 & $13.375$ & $1.0$ & $1.35$ & $1.35$ & $13.23$ & $13.23$ & $1.47$ & $1.47$ & $0.151$ & $0.151$\tabularnewline
\hline
\ttt{DD2-RP14.000} & DD2 & $14.000$ & $1.0$ & $1.35$ & $1.35$ & $13.23$ & $13.23$ & $1.47$ & $1.47$ & $0.151$ & $0.151$\tabularnewline
\hline
\hline
\ttt{DD2-RP11.250-q09} & DD2 & $11.250$ & $0.9$ & $1.35$ & $1.22$ & $13.23$ & $13.19$ & $1.47$ & $1.31$ & $0.151$ & $0.136$\tabularnewline
\hline
\ttt{DD2-RP12.500-q09} & DD2 & $12.500$ & $0.9$ & $1.35$ & $1.22$ & $13.23$ & $13.19$ & $1.47$ & $1.31$ & $0.151$ & $0.136$\tabularnewline
\hline
\ttt{DD2-RP13.000-q09} & DD2 & $13.000$ & $0.9$ & $1.35$ & $1.22$ & $13.23$ & $13.19$ & $1.47$ & $1.31$ & $0.151$ & $0.136$\tabularnewline
\hline
\ttt{DD2-RP14.000-q09} & DD2 & $14.000$ & $0.9$ & $1.35$ & $1.22$ & $13.23$ & $13.19$ & $1.47$ & $1.31$ & $0.151$ & $0.136$\tabularnewline
\hline
\end{tabular}
\caption{Summary of the eccentric binary systems under
  consideration. Listed are the Newtonian periastron distance $r_{p}$, the
  mass ratio $q := M_{2}/M_{1}$, the ADM masses $M_{1,2}$, the TOV
  radii $R_{1,2}$, the baryon masses $M_{b,1,2}$ and the compactness
  ${\cal C}_{1,2} := {M_{1,2}}/{R_{1,2}}$.\label{tab:model-summary}}
\end{table*}

\subsection{Initial data}
\label{sec:initial-data}

We construct our binary neutron-star models as marginally unbound systems
by superposing two boosted nonrotating stellar solutions (TOV). The
initial separation is fixed to $100\,\Msol \approx 150\,\km$ and we
choose the gravitational mass at infinity of the first star to be $M_{1}
= 1.35\,\Msol$, while the mass of the second star is set to be $M_{2} =
qM_{1}$, with $q \leq 1$ the mass ratio. Further fixing the periastron
separation $r_{p}$ and assuming that the stars are on a Newtonian
parabolic orbit, \ie the eccentricity is maximal (${\rm e} = 1$),
determines the initial velocities used for boosting the two TOV
solutions. The properties of the models under consideration are listed in
Table~\ref{tab:model-summary}.

We should note that this initial data is not constraint satisfying, but
approximates the correct initial data as the initial binary separation
goes to infinity. This is why the initial coordinate separation is chosen
to be much larger than for quasi-circular binaries and we make extensive
use of the constraint violation damping provided by the CCZ4
formalism. Confidence on the robustness of our initial data also comes
from similar experience developed when an intrinsic spin angular momentum
was added to binaries in quasi-circular orbits \cite{Kastaun2013}, which
produced results very similar when compared to those obtained with a
consistent calculation of spinning binaries \cite{Bernuzzi2013}. We note
that the approximate nature of our initial data does not cause any
significant GW emission arising from spuriously triggered, (mainly
radial) oscillations in the stars. Such oscillations were observed in an
earlier work \cite{Gold2012}, which was later refined with better initial
data \cite{Moldenhauer2014, Paschalidis2015, East2016, East2016a} that
reproduced the same qualitative physics with respect to the $f$-mode
oscillations. This agreement further supports the point that the initial
data used here, while being only approximate, does incorporate the
essential physics to model these systems.

\subsection{Equations of state}
\label{sec:EOS}

To model the neutron-star matter realistically we employ two different
tabulated nuclear EOSs with full composition and temperature dependence,
namely DD2 \cite{Typel2010} and LS220 \cite{Lattimer91}, where the latter
is used with a nuclear compressibility parameter of $K = 220\,{\rm MeV}$,
which makes it softer than the DD2 EOS. Both of these EOSs can support a
neutron star of mass $\sim 2M_\odot$ as required by pulsar observations
\cite{Demorest2010,Antoniadis2013}, although it should be noted that the
DD2 EOS has a maximum mass which is in slight tension with the most
recent estimates \cite{Margalit2017, Rezzolla2017, Ruiz2017,
  Shibata2017c} and the LS220 EOS predicts stellar radii that are
somewhat larger than what considered to be more likely \cite{Most2018,
  Abbott_etal_2018a}. The combined use of these EOSs serves us to crudely
probe some of the possibilities of composition and temperature-dependent
EOSs, although it would be desirable to consider also softer EOSs, which we
will explore in a future work.

We note that recent work in \cite{Kolomeitsev2016} has found the LS220
EOS (together with other EOSs) to violate conservative constraints on the
symmetry energy, disfavouring this EOS. Nonetheless, we employ this EOS
as a fiducial reference to previous binary neutron-star merger studies,
which have made extensive use of this EOS, especially in the case of
eccentric mergers \cite{Radice2016}. The neutron-star matter is
initially setup in beta equilibrium at the minimal temperature of the EOS
tables ($T=0.01 \, \rm MeV$), where the neutrino chemical potential
$\mu_{\nu}$ vanishes. In turn, this equilibrium fixes the initial
electron (or equivalently proton) fraction $Y_{e}$ distribution in
the stars.

\subsection{Grid setup}

The numerical grid is managed by the mesh-refinement driver \verb+Carpet+
\cite{Schnetter-etal-03b}. It implements a nonuniform grid via a nested
set of movable boxes (box-in-box) together with a hierarchical
(Berger-Oliger-style) timestepping. In these simulations we employ six
refinement levels where each additional level doubles the resolution of
the enclosing one. On the finest grid, which is separately centered
around each neutron star, the resolution in each dimension is $\Delta x =
0.15\, \Msun \approx 220\, \rm{m}$. The physical domain extends to $512\,
\Msun \approx 750\, {\rm km}$, where we track and measure the ejected
material.  We make explicit use of the reflection symmetry relative to
the $z = 0$ plane to reduce the computational cost essentially without
introducing more approximations. The timestep is fixed to $1/6$ of the
grid spacing, corresponding to a Courant-Friedrichs-Lewy (CFL) factor of
$0.15$ and a third-order strong stability preserving Runge-Kutta method
is used for advancing the computation in time.

We note that we have performed a simulation of model \ttt{LS220-RP13.25}
with a higher resolution of $\Delta x = 0.1 \, \Msol$, finding overall
consistency with the results obtained with our reference resolution of
$\Delta x = 0.15 \, \Msol$. More specifically, in the high-resolution
binary we observe that the apoastron separation increases by $4\%$,
showing a slightly different orbital configuration after the first
periastron passage. While the overall waveform shows the same qualitative
features and amplitudes, the ejected mass is increased by approximately
$30\%$. This is not particularly surprising since the dynamics of the
system depends sensitively on the choice made for the impact parameter
$r_{p}$.

We also note that while these studies are far from being a systematic
convergence analysis, they serve our scope. In fact, the variance
introduced because of the finite numerical accuracy -- which causes a
slight drift on the dependence on $r_p$ in the parameter space -- is
compensated by the fact that $r_p$ is treated here as a free parameter
and investigated over a large range. Hence, while the results for a given
value of $r_p$ may not be accurate, the overall dependence of the results
on $r_p$ is.

\subsection{Analysis methods}

\subsubsection{Outflow properties}
\label{sub:outflow}

To measure the properties of the ejected material, we place multiple
spherical detectors at different radii around the origin. As a trade-off
between resolution and distance from the merger remnant, we take the
detector of $200\, \Msun \approx 300\,{\rm km}$ coordinate radius for all
measurements.  Since not all material crossing this surface is unbound,
we have to set a physically motivated threshold. Here we use the so
called ``geodesic criterion'', which is an approximate threshold based on
the Newtonian limit, which gets more accurate far away from the strong
gravity region
\begin{equation}
\label{eq:geodesic_crit}
  u_{t}\le-1\,,
\end{equation}
where $\boldsymbol{u}$ is the fluid four-velocity vector. We note that
this criterion leads in general to a lower estimate of the ejected
mass outflow, which can be a factor of two larger when selecting with a
different criterion, \eg the Bernoulli criterion.  For a discussion of
the impact of this and alternative criteria on the measurements of the
ejected mass, see, \eg Refs. \cite{Kastaun2014, Bovard2016, Bovard2017}.

\subsubsection{Tracers and nuclear-reaction networks}
\label{sub:tracers_nuc}

We employ test-particle tracers which are passively advected along with the
fluid, \ie
\begin{equation}
  \frac{d\boldsymbol{\vec{x}}}{dt}=
  \boldsymbol{\vec{v}}=\alpha\boldsymbol{u}+\boldsymbol{\beta}\,,
\end{equation}
where $\boldsymbol{\vec{x}}$  and $\boldsymbol{\vec{v}}$ are the fluid
element's spatial position and its three-velocity, respectively
\cite{Wanajo2014, Kastaun2016, Mewes2016, Bovard2016}.

Two sets of tracers are initialised throughout the simulations. The first
set of tracers is initialised at the first encounter, where the
separation of the two neutron-star cores is minimal, \ie at
periastron. The second set is initialised once the neutron stars reach
their maximum separation so as to be able to track the material of the
full tidal arms which develop up to the point of merger. The tracers are
placed on the finest refinement level and are sampled uniformly over the
whole density range, since it best reproduces the behaviour of the
ejected matter (see \cite{Bovard2016} for a detailed discussion). Some of
the lowest density regions are excluded to avoid any oversampling at the
boundaries of the finest refinement level. During the evolution the
tracers are updated according to the interpolated values of the fluid
properties at the location of the tracers. As mentioned above, we employ
this two-step procedure to seed tracers because tests have shown that
initialising a set of tracers only at the beginning of the simulation
leads to an underrepresentation of the post-periastron tidal ejecta,
since the majority of tracers initialized near the surface of the neutron
star are ejected during the first periastron passage.

The thermodynamic trajectories of the unbound set of tracers are
subsequently used as input for the \texttt{SkyNet} nuclear-reaction
network \cite{Lippuner2017}. In section \ref{sec:nucleo} we describe
which subset of the unbound tracers is used to compute the total
r-processed material. Such yields are computed using the most recent JINA
REACLIB strong interaction rates \cite{Cyburt2010}, neutron-induced
fission rates with zero emitted neutrons from Ref. \cite{Panov2010} and
spontaneous fission rates with zero emitted neutrons calculated in
Ref. \cite{Roberts2011}.  Furthermore, we use a combination of the rates
from \cite{Fuller82} and JINA REACLIB \cite{Cyburt2010}
for the weak interactions.

The tracer trajectories are trimmed to begin once their temperatures
drop below $8 \times 10^{6}\,{\rm K}$ and end before their densities
drop to near-atmosphere values, or if the fluid element reaches the outer
boundary. From there on, we assume a spherical, free expansion of the
material, leading to the subsequent extrapolations beginning after the
last available data of the tracer trajectory \cite{Fujimoto2008,
  Korobkin2012}
\begin{align}
  r(t) &= r_0 + v_0t\,,
  \label{eq:expansion-r}\\
  \rho(t) &= \rho_0 \left(\frac{t}{t_0}\right)^{-3}\,,
  \label{eq:expansion-hor}\\
  T(t) &= T\left[s,\rho(t),\Ye(t)\right]\,.
  \label{eq:expansion-T}
\end{align}
where the temperature is computed from the modified Helmholtz EOS
implemented in \texttt{SkyNet} (see \cite{Timmes1999, Timmes2000,
  Lippuner2017} for details). The temperature threshold for the
transition from nuclear statistical equilibrium (NSE) to the full nuclear
network is assumed to be at $T = 7\, \rm GK$ and the r-process is
followed up to $\sim 3\times 10^{6}\, {\rm yr}$ after merger. This setup
is rather standard and similar to the one employed in
Ref. \cite{Bovard2017}, where a different nuclear-reaction network code
(\texttt{WinNet}) was used.

\subsubsection{Gravitational-wave signals}
\label{sub:grav-waves-theo}

The GW strain is extracted using the standard Newman-Penrose formalism
\cite{Newman62a}, in which a particular contraction of the Weyl tensor
with a suitably chosen null tetrad yields a gauge-invariant scalar
quantity $\psi_{4}$ that encodes the outgoing gravitational radiation
(see \eg \cite{Bishop2016}). More specifically, $\psi_{4}$ is related to
the second time derivative of the two strain polarisations
$\ddot{h}_{+,\times}$ by

\begin{equation}
  \ddot{h}_{+} - {\rm i} \ddot{h}_{\times} = \psi_4
  = \sum_{\ell=2}^{\infty} \sum_{m=-\ell}^{\ell}
  \psi_4^{\ell m}\;_{-2}Y_{\ell m}(\theta,\varphi)\,,
\end{equation}
where we introduced also the multipole expansion of $\psi_{4}$ in
spin-weighted spherical harmonics \cite{Goldberg:1967} of spin-weight
$s=-2$. In addition, we compute the radiated energy and angular momentum
from the radiated torque as
\begin{equation}
\dot{J}=\frac{r^{2}}{16\pi}\Im\left(\sum_{\ell=2}^{\infty}
\sum_{m=-\ell}^{\ell}m\int_{-\infty}^{t}\int_{-\infty}^{t'}\bar{\psi}_{4}^{\ell
  m}dt''dt' \int_{-\infty}^{t}\psi_{4}^{\ell m}dt'\right)\,,
\label{eq:gw-torque}
\end{equation}
and the radiated power as
\begin{equation}
\dot{E}=\frac{1}{16\pi}\sum_{\ell=2}^{\infty}\sum_{m=-\ell}^{\ell}
\int_{-\infty}^{t}\left|r\psi_{4}^{\ell m}dt'\right|^{2}\,.
\label{eq:gw-power}
\end{equation}

The double integrations in time introduce substantial nonlinear drifts of
the strain as well as of the total radiated angular momentum and
energy. To eliminate this artificial drift coming from random numerical
noise in the $\psi_{4}$ data, we perform all time integrations using the
fixed-frequency integration from Ref. \cite{Reisswig:2011}.

We define the power spectral density (PSD) of the effective strain
amplitude as
\begin{equation}
\tilde{h}\left(f\right):=
\sqrt{\frac{|\tilde{h}_{+}|^{2}+|\tilde{h}_{\times}|^{2}}{2}},
\label{eq:eff-psd}
\end{equation}
with
\begin{equation}
\tilde{h}\left(f\right)_{+,\times}=\begin{cases}
\int h_{+,\times}\left(t\right)e^{-i 2 \pi f t}dt & f\geq0\\
0 & f<0
\end{cases}.
\label{eq:eff-psd-ft}
\end{equation}

Note, that the PSD is most informative for signals with well-separated
features or with secular variations in frequency, such as for
quasi-circular sources. On the other hand, for eccentric sources the
instantaneous GW frequency is not a monotonic function of time and signal
power in a certain frequency range can overlap in frequency space despite
being well-separated in time. The GW spectrograms (\cf
Fig. \ref{fig:gws-combined}) are a much more faithful representation of
how the signal manifests itself in the detector (for either
quasi-circular or eccentric sources).

\begin{figure}[h]
\begin{center}
\includegraphics[width=1.0\columnwidth]{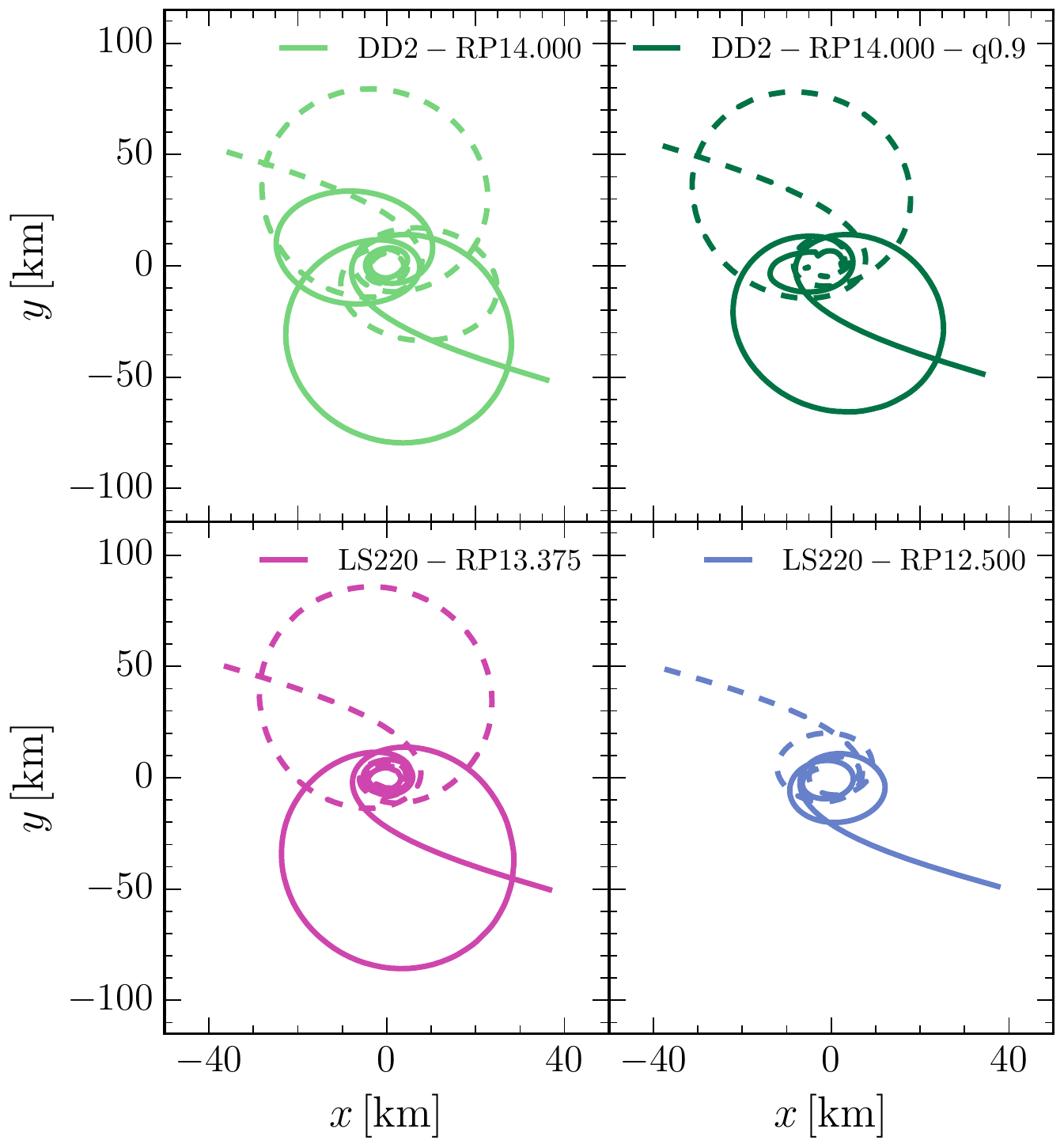}
\caption{Orbital trajectories of the barycenters of the neutron stars for
  four representative models, spanning a large range in periastron values, mass
  ratio and EOS. Note that some binaries have multiple periastron
  approaches (\eg \ttt{DD2-RP14.000}) and that solid/dashed lines refer
  to the trajectories of either member of the binary.}
\label{fig:orbits}
\end{center}
\end{figure}

\section{Results}
\label{sec:results}

\subsection{Orbital dynamics}
\label{sub:orbit-dynamics}

Highly eccentric binary mergers exhibit richer dynamics and
phenomenology when compared to quasi-circular mergers and more parameters
are needed to fully describe the orbital dynamics in its general
form. This is shown in Fig. \ref{fig:orbits}, which reports the orbital
trajectories of the barycenters\footnote{Note that we here adopt the
  Newtonian definition of barycenter of a rest-mass density
  distribution.} of the neutron stars for four representative models,
spanning the set considered in initial periastron distance, mass ratio
and EOS. Note that some binaries in our set have multiple (up to three)
periastron approaches, \ie \ttt{DD2-RP14.000}; furthermore, the location
of the apoastron can be very large for ``bare'' captures, \ie
\ttt{LS220-RP13.375}, or very small \ie \ttt{DD2-RP12.500}.

\begin{figure*}
\begin{center}
\includegraphics[width=1.5\columnwidth]{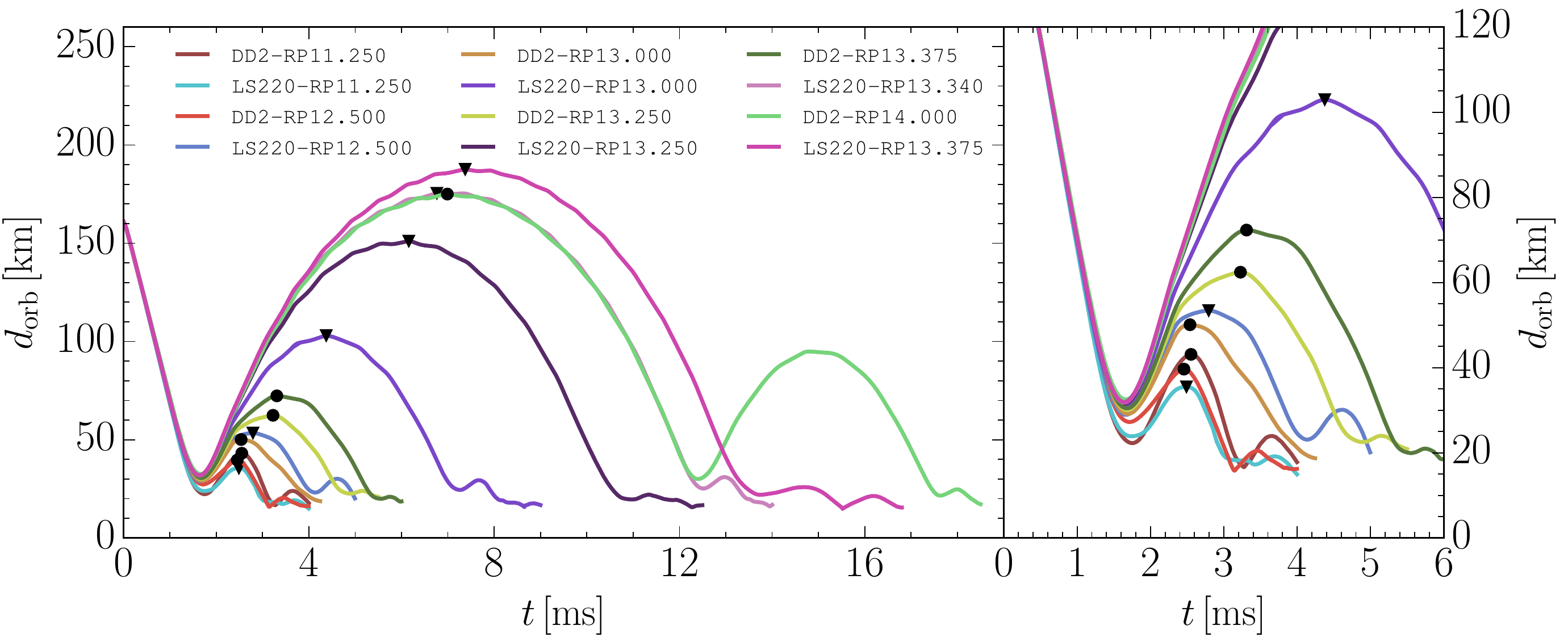}
\caption{Proper separation $d_{\rm orb}$ of the stellar barycenters
  for the equal-mass models as a function of time. The symbols mark
  the maximal separations after the first periastron passage. Note
  that the curves are truncated at the moment of merger.}
\label{fig:orb-sep}
\end{center}
\end{figure*}
\begin{figure*}[t]
\begin{center}
\includegraphics[width=1.0\textwidth]{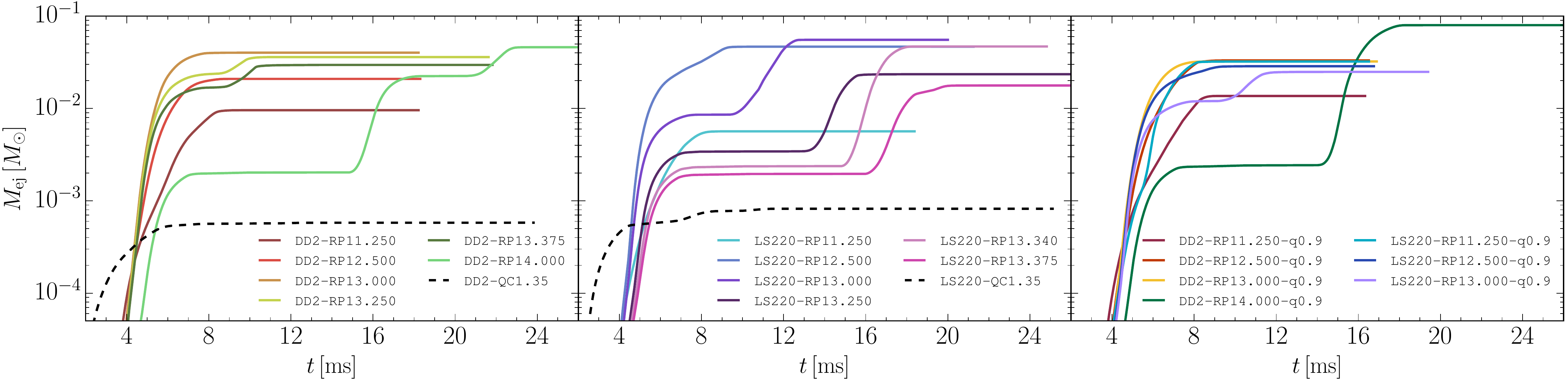}
\caption{\textit{Left panel:} Dynamically ejected mass $M_{\rm ej}$ at a
  $r=200\,\Msol$ detector as a function of time for both EOS and the
  different periastron values listed in
  Table~\ref{tab:model-summary}. For comparison the same is shown for a
  quasi-circular binary of the same mass and EOS from
  \cite{Bovard2017}. Note that during a single periastron passage alone,
  more ejecta compared to an entire quasi-circular inspiral are
  produced. \textit{Middle panel:} the same as the left panel but for the
  LS220 EOS. \textit{Right panel:} the same as the left panel but for
  unequal-mass binaries with $q=0.9$.}
\label{fig:massej-equal}
\end{center}
\end{figure*}

The systems under consideration here are all setup in a marginally bound
orbit with maximum Newtonian eccentricity ${\rm e} = 1$, and with varying
periastron separation $r_p$. This restriction to a one-dimensional
parameter space expressed in terms of a Newtonian description allows us
to probe tidal interactions of vastly different strength that are similar
to dynamical-capture scenarios that can occur in the cores of globular
clusters \cite{Lee2010,Thompson2011} or in stellar cusps in galactic
cores \cite{Oleary2009,Kocsis2012,Antonini2012}. Note however that
because of GW losses, all orbits in this study are actually bound, either
in the sense that they merge on the first encounter or lose sufficient
orbital energy through tides and GW emission that they move apart only up
to a finite apoastron distance before returning for the next encounter.

These post-capture apoastron distances increase monotonically with
increasing periastron separation because gravitational radiation and the
tidal interactions become weaker. This is shown in
Fig.~\ref{fig:orb-sep}, where we report the proper separation between the
barycenters, $d_{\rm orb}$, for several of the binaries considered
here. By comparing cases with the same initial $r_p$ but different EOSs,
it is possible to see clearly that the EOS leads to substantially
different orbital dynamics. These differences arise from the interactions
of the stars with the binary tidal field, which depends on the EOS. From
Fig.~\ref{fig:orb-sep} it is also possible to deduce that as the initial
impact parameter $r_p$ is increased, the separation after the first
periastron passage increases, with binaries that either merge with the
second periastron passage (this is the large majority of the cases
considered here), or that fly off for yet another orbit. This is indeed
the case of the binary \ttt{DD2-RP14.000}, which has a periastron
parameter of $r_{p} = 14.0\,\Msol$ and undergoes two close passages and
merges at the third. Overall, the examples collected in
Fig.~\ref{fig:orb-sep} clearly show that the orbital dynamics for our
binaries is only very poorly described by a simplified test-particle
approximation, be it in Newtonian gravity or in terms of a geodesic
motion in a stationary spacetime in general relativity. In practice,
finite-size effects, tidal interactions and GW losses make the dynamics
of the system far more complex and accurately described only in terms of
fully nonlinear numerical simulations.

We should note that although the pre-merger evolution of the binary
\ttt{DD2-RP14.000} in Fig.~\ref{fig:orb-sep} extends only to a time $\sim
17\,{\rm ms}$, the computational costs associated with this binary are
indeed extremely high. This is because the grid structure that needs to
be employed for a proper analysis of the two stars as they move away from
each other requires high-resolution refinement boxes that are very
extended and hence computationally expensive. As a reference, the
computational costs of the binary \ttt{DD2-RP14.000} for a timescale
of $17\,{\rm ms}$ would correspond to a post-merger evolution of $\sim
25\,{\rm ms}$ for the same binary merging from a quasi-circular
orbit.

Note that all remnants formed in this study are HMNSs, \ie their mass
exceeds the maximum mass of a uniformly rotating neutron star supported
by the given EOS $M_{\rm max}$, which is then related in a
quasi-universal manner to the maximum mass of a nonrotating
configuration, $M_{_{\rm TOV}} \simeq 1.2 M_{\rm max}$
\cite{Breu2016}\footnote{Given a law of differential rotation, a
  universal relation can be found also by the maximum mass supported
  through differential rotation and the maximum mass of a nonrotating
  configuration \cite{Weih2017}.}. The merged object, instead, is in a
metastable equilibrium supported by differential rotation, which has a
quasi-universal rotation profile in case of quasi-circular merger
remnants \cite{Hanauske2016}. We evolve the simulations until the mass
density flux at the detector placed at a distance of $200 \,\Msol\approx
300\,{\rm km}$ becomes essentially zero. During this time, none of the
merged objects has yet collapsed to a black hole.

\begin{table*}[t]
\begin{tabular}{c|c|c|c|c|c|c|c}
Model & $M_{\rm ej}\,\left[10^{-2}M_{\odot}\right]$  & $\left\langle Y_{e}\right\rangle $ & $\left\langle s\right\rangle \,\left[{\kB}/{\rm baryon}\right]$ & $\left\langle v_{\rm ej}\right\rangle \,\left[c\right]$ & $\left\langle v_{\infty}\right\rangle \,\left[c\right]$ & $J_{\rm rad}\,\left[M_{\odot}^{2}\right]$ & $E_{\rm rad}\,\left[10^{-2}M_{\odot}\right]$\tabularnewline
\hline
\hline
\ttt{LS220-RP11.250}    & $0.57$ & $0.07$ & $7.6$ & $0.22$ & $0.15$ & $0.67$ & $2.20$\tabularnewline
\hline
\ttt{LS220-RP12.500}    & $4.67$ & $0.06$ & $4.6$ & $0.24$ & $0.18$ & $1.21$ & $4.07$\tabularnewline
\hline
\ttt{LS220-RP13.000}    & $5.54$ & $0.07$ & $5.0$ & $0.23$ & $0.16$ & $1.22$ & $4.05$\tabularnewline
\hline
\ttt{LS220-RP13.250}    & $2.35$ & $0.07$ & $6.3$ & $0.22$ & $0.15$ & $1.16$ & $4.31$\tabularnewline
\hline
\ttt{LS220-RP13.340}    & $4.69$ & $0.07$ & $5.7$ & $0.23$ & $0.16$ & $1.20$ & $4.04$\tabularnewline
\hline
\ttt{LS220-RP13.375}    & $1.77$ & $0.09$ & $8.1$ & $0.21$ & $0.13$ & $1.61$ & $5.86$\tabularnewline
\hline
\hline
\ttt{LS220-RP11.250-q09} & $3.22$ & $0.04$ & $2.3$ & $0.22$ & $0.15$ & $0.65$ & $2.13$\tabularnewline
\hline
\ttt{LS220-RP12.500-q09} & $2.87$ & $0.05$ & $4.1$ & $0.24$ & $0.18$ & $0.53$ & $1.89$\tabularnewline
\hline
\ttt{LS220-RP13.000-q09} & $2.49$ & $0.07$ & $6.0$ & $0.23$ & $0.17$ & $0.94$ & $3.27$\tabularnewline
\hline
\hline
\ttt{DD2-RP11.250}       & $0.96$ & $0.12$ & $8.9$ & $0.23$ & $0.16$ & $0.93$ & $2.94$\tabularnewline
\hline
\ttt{DD2-RP12.500}       & $2.09$ & $0.05$ & $4.4$ & $0.23$ & $0.16$ & $0.78$ & $2.42$\tabularnewline
\hline
\ttt{DD2-RP13.000}       & $4.02$ & $0.05$ & $3.2$ & $0.26$ & $0.20$ & $0.75$ & $2.60$\tabularnewline
\hline
\ttt{DD2-RP13.250}       & $3.60$ & $0.06$ & $4.9$ & $0.25$ & $0.19$ & $0.80$ & $2.39$\tabularnewline
\hline
\ttt{DD2-RP13.375}       & $2.96$ & $0.06$ & $5.2$ & $0.24$ & $0.18$ & $0.76$ & $2.37$\tabularnewline
\hline
\ttt{DD2-RP14.000}       & $4.61$ & $0.10$ & $7.2$ & $0.21$ & $0.13$ & $1.04$ & $3.21$\tabularnewline
\hline
\hline
\ttt{DD2-RP11.250-q09}   & $1.37$ & $0.09$ & $7.5$ & $0.21$ & $0.14$ & $0.53$ & $1.59$\tabularnewline
\hline
\ttt{DD2-RP12.500-q09}   & $3.31$ & $0.05$ & $4.2$ & $0.22$ & $0.15$ & $0.41$ & $1.20$\tabularnewline
\hline
\ttt{DD2-RP13.000-q09}   & $3.23$ & $0.05$ & $3.3$ & $0.25$ & $0.19$ & $0.72$ & $2.51$\tabularnewline
\hline
\ttt{DD2-RP14.000-q09}   & $7.99$ & $0.06$ & $2.8$ & $0.22$ & $0.15$ & $0.45$ & $1.33$\tabularnewline
\hline
\end{tabular}
\caption{Summary of the results for the eccentric binary systems under
  consideration. Listed are the dynamically ejected mass $M_{\rm ej}$,
  the mass-averaged electron fraction $Y_{e}$, specific entropy $s$, the
  mass-averaged ejecta velocity $\langle v_{\rm ej} \rangle $ and
  asymptotic ejecta velocity $\langle v_{\infty} \rangle$ as well as the
  angular momentum $J_{\rm rad}$ and energy $E_{\rm rad}$ radiated by
  GWs. \label{tab:model-results}}
\end{table*}

\subsection{Mass ejection}
\label{sec:me}

We next discuss the properties of the ejected mass of all simulations
shown in Table~\ref{tab:model-summary}. The amount of unbound mass and
its properties in terms of electron (proton) fraction, specific entropy
and velocity are analysed as a function of the periastron separation
$r_{p}$ and the most important results are summarised in
Table~\ref{tab:model-results}.

As already discussed in Sec. \ref{sub:outflow}, we analyze the ejected
mass using a spherical detector at a coordinate radius of $200 \, \Msun
\approx 300\,{\rm km}$. The total amount of unbound material is computed
by integrating the rest-mass density flux of every fluid element on the
sphere in time, as long as it obeys the geodesic criterion. The total
amount of ejected mass is then given by integrating the mass flux over
the whole sphere. Together with the mass flux we also compute the
electron (proton) fraction, the specific entropy and the velocity of the
unbound fluid elements.  We should note that we could have used detectors
at larger distances than the reference one of $200 \, \Msun \approx
300\,{\rm km}$. While doing so would increase the amount of measured
ejected mass (which can increase up to $30\%$ when going to a radius of
$400 \, \Msun$), such measurement would also be quite inaccurate. This is
because of two distinct reasons, both numerical and physical. First, as the
ejected matter reaches more distant detectors its rest-mass density drops
near the atmosphere level of the simulation, which, we recall, is already
11 orders of magnitude smaller than the maximum rest-mass density in the
simulations; in these regimes the numerical truncation error is clearly
largest. Second, as the fluid expands and rarefies, the assumption of NSE
used to construct the EOS tables is no longer correct, thus invalidating
the analysis of its thermodynamic properties in these low-density
regimes. In view of these considerations, we have preferred to use the
measurements made on a detector which is not the largest one but provides
us with more accurate results. Finally, as already stated in
Sec.~\ref{sub:outflow}, the largest source of uncertainty in the total
ejected mass depends is given by the choice of unbound-matter criterion
[\eg Eq. \eqref{eq:geodesic_crit}] rather than by the detector radius.

\begin{figure}
\begin{center}
\includegraphics[width=1.0\columnwidth]{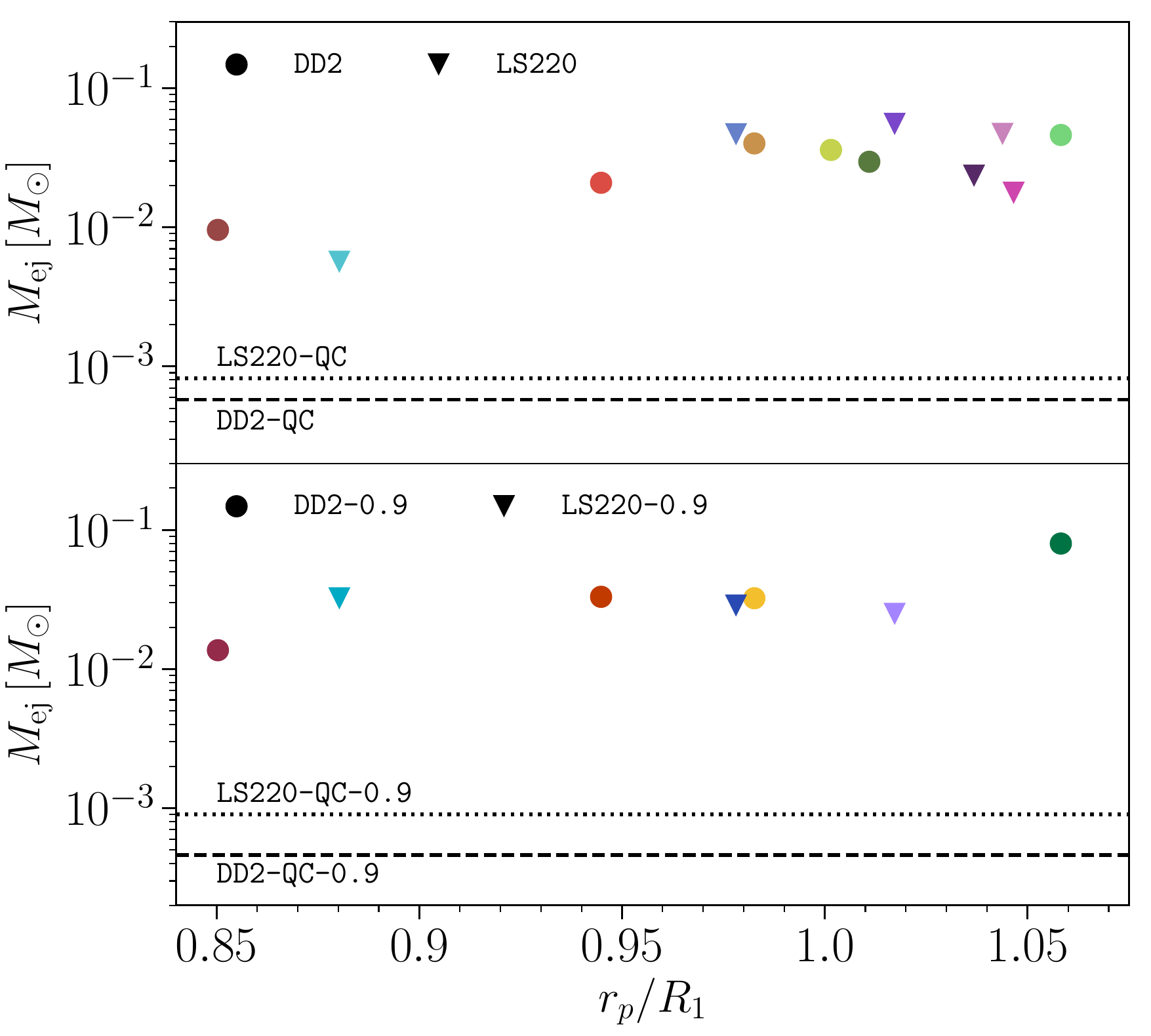}
\caption{Dynamically ejected mass $M_{\rm ej}$ at a $r=200\,\Msol$ detector
  as a function of the dimensionless quantity ${r_{p}}/{R_{1}}$ for both
  EOS and mass ratios, where $R_{1}$ is the TOV radius of the more
  massive companion.  For comparison the same is shown for equal-mass
  quasi-circular binaries of the same mass and EOS from
  \cite{Bovard2017}.}
\label{fig:massej-all}
\end{center}
\end{figure}

The dynamically ejected masses for the eccentric equal-mass binaries are
shown in Fig.~\ref{fig:massej-equal} for the two EOSs considered (see
legend) and are compared with the corresponding ejected mass in the
quasi-circular case for the DD2 EOS (dashed black line). It is quite
clear that every single periastron passage in our models produces more
ejecta than an entire quasi-circular merger, excluding longterm
post-merger evolution. Overall, the models show very similar behaviour
with cumulative outflows at every close encounter which last a few
milliseconds.  Note that shock-heating during merger does create
additional amounts of hot unbound material, but that this is only a small
contribution when compared to the cold material ejected because of tidal
interactions.
Similar to the results in Refs. \cite{Radice2016} for the LS220 EOS, the
eccentric mergers considered in this work can lead to almost two orders
of magnitude larger ejected masses when compared to quasi-circular
mergers (in this case we compare to models of \cite{Bovard2017} with the
same mass and EOS).

The dependence of the ejected mass on the EOS, on the impact parameter
and on the mass ratio of the binary can be appreciated from
Fig.~\ref{fig:massej-all}. First of all, the total amount of ejected mass
is robustly larger than the corresponding quasi-circular binaries,
despite the differences in the orbital parametrization, as well as the
EOS and the mass ratio. Additionally, there is no definitive trend
of larger mass ejection for unequal-mass binaries observed in
quasi-circular binary neutron-star mergers, even though the lower mass
companion is tidally disrupted at the second encounter. Note that for the
unequal-mass models \ttt{LS220-RP11.250-q09}, \ttt{DD2-RP11.250-q09}, and
\ttt{DD2-RP12.500-q09}, the ejected mass is larger than in corresponding
equal-mass models. However, a further increase in $r_{p}$ leads to the
opposite behaviour, only \ttt{DD2-RP14.000-q09} giving again a larger amount.

Figure \ref{fig:massej-all} suggests the presence of a non-monotonic
behaviour of the ejected mass with $r_p$ and which has a rather natural
interpretation: for small values of $r_p/R_1$ one expects more ejected
mass than in an equivalent binary in a quasi-circular orbit, while in the
limit of very large values of $r_p/R_1$ one expects to recover the amount
obtained in the quasi-circular case. What happens in between these two
limits is far more complex and depends also on the matching between the
oscillations of the two stars and the actual merger. The appearance of
additional local maxima is possible and there could even be an evidence
in the data, but this evidence is only marginal.

Note also that in each of the three passages in models
\ttt{LS220-RP13.340} and \ttt{LS220-RP14.000}, the tidal interactions
lead to subsequent large releases of unbound material. The additional
encouter in model \ttt{LS220-RP13.340} is strongly suppressed due to the
outer layers of both stars overlapping throughout the second encounter,
leaving only the cores of both stars to separate slightly again before
eventually merging.

Going beyond a periastron separation of ${r_{p}}/{R_{1}} = 1.07$, as done
for model \ttt{LS220-RP13.375}, the system merges again at the second
encounter and thus produces less ejecta. Furthermore, when considering
less eccentric orbits, multiple encounters do not only occur more easily,
but are in fact unavoidable, see \eg
Refs. \cite{Gold2012,Gold:2012tk}. For ${r_{p}}/{R_{1}} > 1.1$ we expect
a recurrence of this pattern for fine-tuned initial orbital
configurations, although exploring the space of parameters to this level
of detail, possibly for critical behaviour, is beyond the scope of this
paper.

\begin{figure*}[]
\begin{center}
\includegraphics[width=1.9\columnwidth]{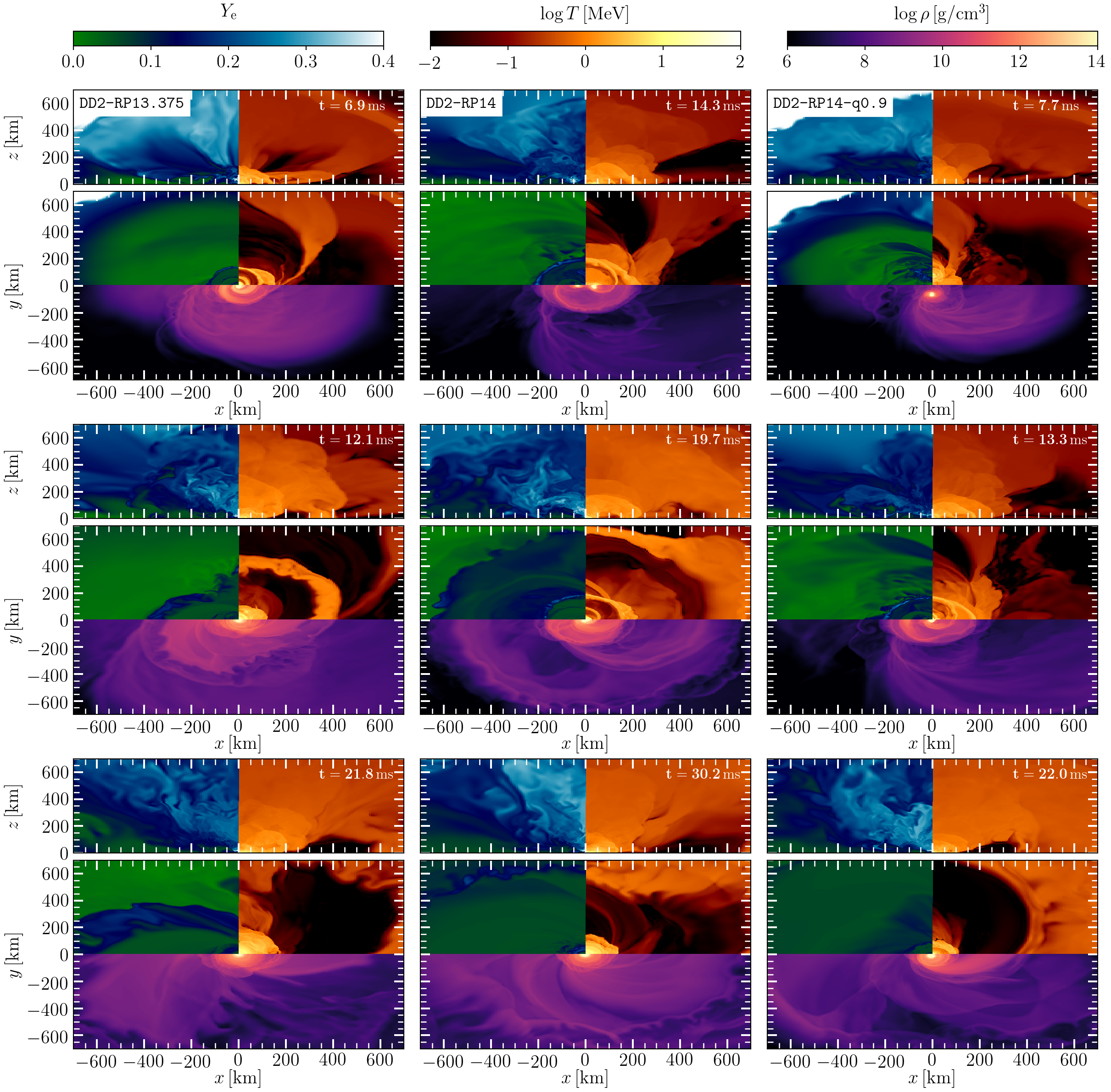}
\caption{Snapshots of the electron fraction $Y_e$, logarithmic
  temperature $T$ and rest-mass density $\rho$ in the equatorial
  $(x,y)$ and meridional $(x,z)$ plane for three models with DD2 EOS.
  Due to the different orbital dynamics different times are shown.
  For the \ttt{DD2-RP13.375} and \ttt{DD2-RP14.000-q09} models we
  show snapshots after the first encounter with the first ejecta (top)
  followed by the merger (middle) and ultimately the remnant when all
  ejecta crossed the $200\,\Msol$ detector (bottom). The same is shown
  for \ttt{DD2-RP14.000} but starting with the second encounter.}
\label{fig:ye-temp-overview}
\end{center}
\end{figure*}

In summary: the typical amount of ejected material is in the range of a
few $10^{-2}\,\Msol$ up to almost $10^{-1}\,\Msol$. As already
conjectured in \cite{Radice2016}, the latter is expected to constitute an
upper limit also for orbits undergoing more than two encounters. Since
the system loses orbital energy and angular momentum with each close
passage, the orbit becomes less eccentric and the amount of ejecta should
approach that of quasi-circular binaries in the last few close
passages. Moreover, we find that the models with multiple encounters lead
to the largest amounts of unbound mass, thus providing a confirmation of
the results found in \cite{Radice2016,East2012c}. The largest amount of
ejecta were produced by the \ttt{DD2-RP14.000-q09} binary, which shows
both a partial third encounter and tidal disruption of the lower mass
companion.

\subsection{Properties of the dynamical ejecta}
\label{sub:dynejec-props}

\begin{figure*}
  \begin{center}
    \includegraphics[width=1.0\textwidth]{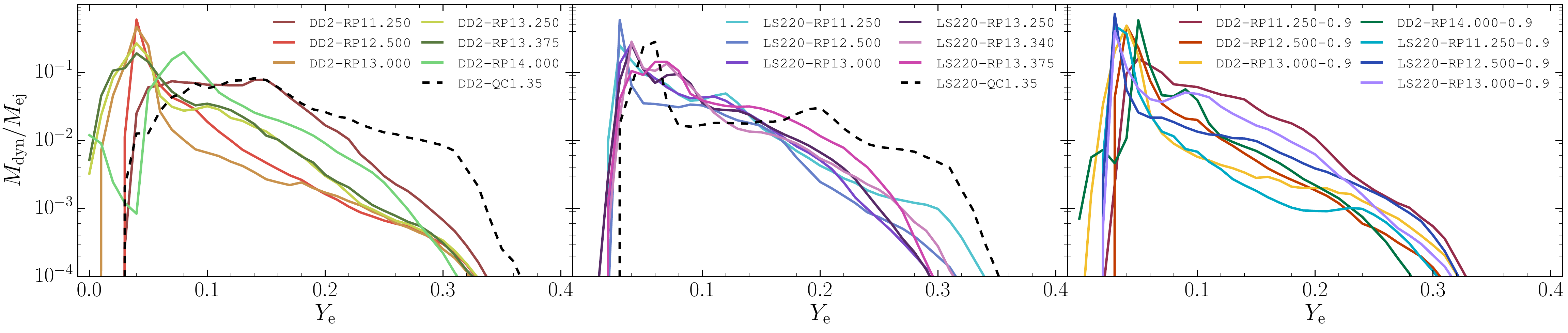}
    \caption{\textit{Left panel:} $Y_{e}$ distribution of the dynamical
      ejecta masses $M_{\rm dyn}$ normalized to the total ejected mass
      $M_{\rm ej}$ at a $r=200\,\Msol$ detector for the equal-mass
      binaries listed in Table~\ref{tab:model-summary}. For comparison
      the same is shown for a quasi-circular binary of the same mass and
      EOS from \cite{Bovard2017}. \textit{Middle panel:} the same as in
      the left panel but for the LS220 EOS. \textit{Right panel:} the
      same as in the left panel but for the $q = 0.9$ binaries listed in
      Table~\ref{tab:model-summary}.}
    \label{fig:ejecta-ye}
  \end{center}
\end{figure*}

In what follows we discuss the physical properties of the dynamical
ejecta, namely: the average electron fraction $Y_{e}$, the specific
entropy $s$, and the velocity $v_{\rm ej}$ of the ejecta; all of these
properties are summarised in Table~\ref{tab:model-results} for the
binaries considered here.

A good general overview of the dynamics and properties of the (dynamical)
ejecta is given in Fig.~\ref{fig:ye-temp-overview}, which shows the
equatorial [$(x,y)$ plane] and meridional [$(x,z)$ plane] slices through
the simulation domain in terms of the electron fraction $Y_{e}$, the
rest-mass density $\rho$, and the temperature $T$ (see different
colourcodes). The latter is directly related to the entropy per baryon
$s$ and together with $Y_{e}$, they represent one of the most important
quantities for the r-process nucleosynthesis taking place in the
dynamical ejecta. More specifically, we show in the three columns of
Fig.~\ref{fig:ye-temp-overview} the following binaries: a representative
case of the DD2 EOS, \ttt{DD2-RP13.375}, which refers to a model with one
periastron passage followed by the merger, the unequal-mass binary
\ttt{DD2-RP14.000-q09}, which refers to the model with the largest amount
of ejected material, and to the binary \ttt{DD2-RP14.000}, which
represents a binary with three periastron passages and includes the
merger. The three rows, instead, illustrate three different times
throughout the simulation. The first shows the binary after the first
periastron passage, the second row refers instead to a time after merger,
and the third row shows when almost all the ejecta matter has crossed the
$200\,\Msol$ detector. Note, that for the \ttt{DD2-RP14.000} binary the
last passage before merger is the second, so we start from the second
encounter.

In order to be efficient, the r-process requires a very neutron-rich
environment from which neutrons can be accessed; hence, the electron
fraction $Y_{e}$ can be used to effectively describe the availability of
such neutrons in the matter ejected during and after the merger. As
described in Sec.~\ref{sec:grhydro}, we model weak interactions by a
leakage scheme, so that the electron fraction $Y_{e}$ is not simply
advected with the fluid, but can and does change due to electron and
positron captures, in particular in the shock-heated material.  In
practice, a large value of the electron fraction is therefore a proxy for
a relatively ``neutron-poor'' (electron rich) material, which will
therefore lead to a suppressed r-process.

As can be inferred in Fig.~\ref{fig:ye-temp-overview}, there is a clear
difference in $Y_{e}$ between the ejecta in the equatorial plane and the
ejecta off the plane and, indeed, the distribution in $Y_{e}$ can be
associated to two different components. The first one comes from the cold
neutron-rich material that is ejected by the tidal torques, while the
second one originates from the shock-heated, high-temperature
material. Both of these components can be recognised in
Fig~\ref{fig:ye-temp-overview} as the equatorial and the polar components
of the ejecta, respectively. Note, that due to both the higher electron
fraction and the much lower density, the polar component of the ejected
matter is a poor site for the heaviest r-process elements, in contrast to
the matter ejected on the equatorial plane by tidal torques.

The distributions of $Y_{e}$ in the dynamical ejecta for the equal-mass
models are shown in Fig.~\ref{fig:ejecta-ye}, which also reports the
corresponding distribution of a quasi-circular model of the same mass and
EOS of \cite{Bovard2017} (dashed black lines). Note that the qualitative
features of the distributions are very similar for the different
models. This conclusion is slightly different from that reported in
Ref. \cite{Radice2016} and it is possible that the use of tracer
particles can account for these small discrepancies.

An obviously prominent feature in the $Y_e$ distributions in
Fig.~\ref{fig:ejecta-ye} are the peaks at very low values of the electron
fraction, \ie $Y_{e} \in [0.04,0.08]$, which are followed by a sharp drop
to higher values of $Y_{e} \lesssim 0.35$. More importantly, the
eccentric binaries are not able to reach the high $Y_{e}$ tails that is
instead achieved in the corresponding quasi-circular models. The only
exception to this behaviour is given by the \ttt{DD2-RP11.250} model,
which shows a distribution much closer to the quasi-circular model,
without showing the pronounced low $Y_{e}$ peak of the other models. The
reason behind this behaviour is that this model is undergoing a direct
and violent merger without an additional periastron passage it is more
similar to the quasi-circular case.

\begin{figure*}
  \begin{center}
    \includegraphics[width=1.0\textwidth]{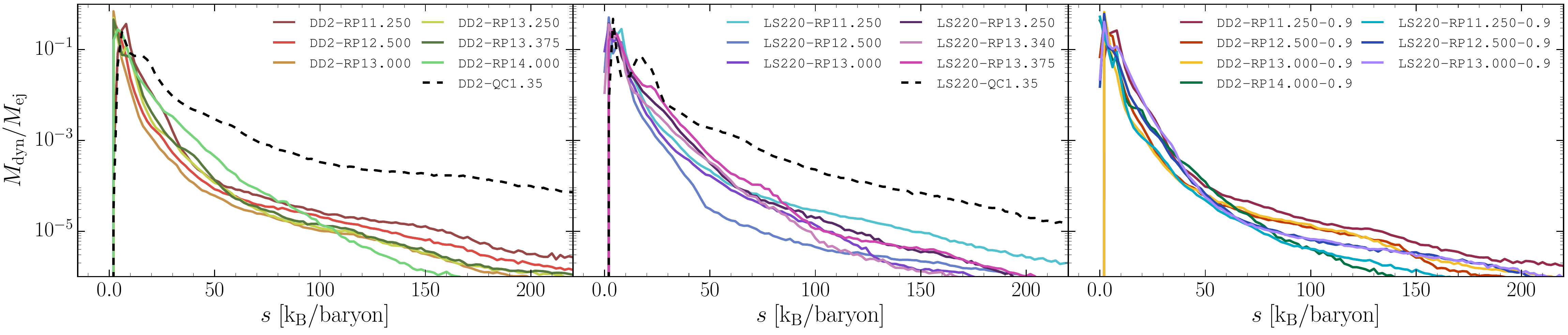}
    \vskip 0.25cm
    \includegraphics[width=1.0\textwidth]{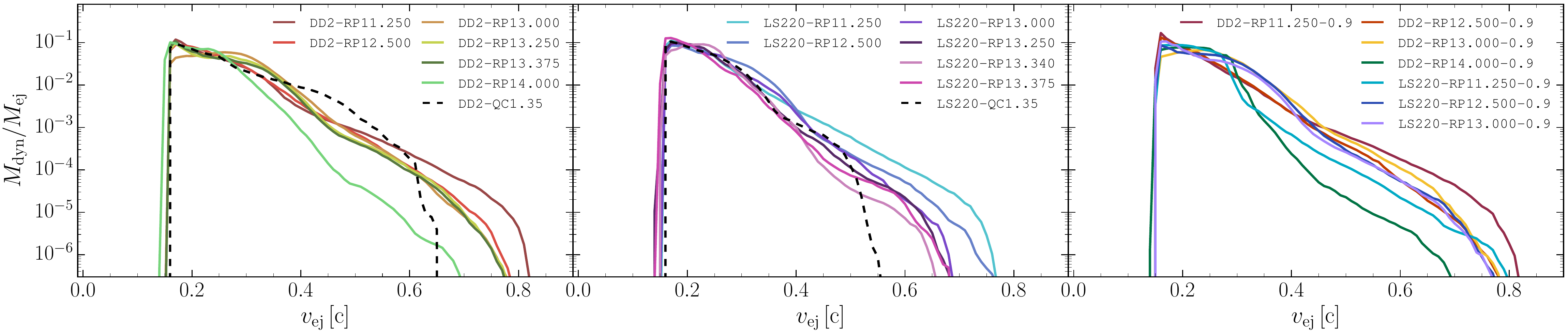}
    \caption{The same as in Fig. \ref{fig:ejecta-ye}, but for the
      distribution of the specific entropy (top panels) and for the
      distribution of the velocity of the ejected material (bottom
      panels).}
    \label{fig:ejecta-entropy}
  \end{center}
\end{figure*}

All in all, it is clear that the LS220 models are closer to the
quasi-circular distribution, but with more material exhibiting lower
$Y_{e}$ values. Whether or not this can be explained purely by different
stiffness, which would also be applicable to other EOS, cannot be
answered with our limited data set, but appears reasonable on the basis
of our simulations. For increasing periastron separation, the peak in the
distributions is shifted to higher $Y_{e}$ but stays well below
$0.1$. Comparing this behaviour to that of the DD2 models -- with the
exception of the special \ttt{DD2-RP11.250} binary -- this shift is
absent and all models peak at $Y_{e} \simeq 0.04$. Leaving aside the
differences across the various EOSs, it is apparent that neither the EOS
nor the periastron distance have a significant impact on the $Y_{e}$
distribution, as long as the binary does not undergo a direct or almost
head-on merger.

When considering instead the unequal-mass models, which are shown in the
top-right panel of Fig.~\ref{fig:ejecta-ye}, the qualitative behaviour
does not change considerably. The distributions feature again a peak at
$Y_{e} \lesssim 0.06$ and a sharp decline to higher values. The dynamical
ejecta are dominated by extremely neutron-rich material, which is also
clearly visible from the average $Y_{e}$ of each model listed in
Table~\ref{tab:model-results}, where $\left\langle Y_{e}\right\rangle
\lesssim 0.1$ for all the models.

What is also clear from Fig.~\ref{fig:ejecta-ye} is that -- for a fixed EOS
and mass ratio -- there is a trend to produce distributions with
systematically larger values of $Y_{e}$ as the impact parameter is
increased. This behaviour is not surprising as it is possible to consider
the quasi-circular binaries (which reveal the largest electron-fractions)
as a limiting case of eccentric binaries with very large impact
parameter. Also, as can be seen from Fig.~\ref{fig:massej-equal}, the
amount of ejected material is dominated by the second encounter for the
models with ${r_{p}}/{R_{1}} \apprge 1.03$, so that the binary has
already lost a large part of its initial eccentricity and the orbit is
thus much closer to a quasi-circular one.

The considerations made above, while robust, neglect what could be an
important effect on the $Y_{e}$ distribution, namely, the accounting for
neutrino absorption in the ejecta. This process, which can be modeled by
a more sophisticated radiative-transfer scheme (\eg an M1-scheme) has
been shown that when taken into account in simulations of compact
binaries in quasi-circular orbits, \eg in Refs. \cite{Foucart2015,
  Foucart2015b, Sekiguchi2015, Foucart2016a, Sekiguchi2016}, can lead to
overall higher values of $Y_{e}$. Note, that both EOSs under
consideration lead to rather large stellar radii. While we expect the
same qualitative behaviour for slightly smaller impact parameters, a
softer EOS could lead to larger high-$Y_{e}$ tails.  We leave the
investigation of the impact of neutrino absorption and of softer EOSs on
eccentric binary merger ejecta to future studies.

The second important thermodynamic quantity for the r-process is the
entropy per baryon, whose distribution is shown in the top panels of
Fig.~\ref{fig:ejecta-entropy}. This quantity is intimately connected to
the shock-heated matter in the ejecta, so that the material of higher
temperature visible in Fig.~\ref{fig:ye-temp-overview} effectively trace
material with larger entropy. In turn, the entropy produced in these
shocks impacts the neutron-to-seed ratio of the material and has
therefore a direct influence on the r-process and the nucleosynthesis.
As can be seen in Fig.~\ref{fig:ye-temp-overview}, most of the equatorial
ejecta are of low temperature when compared to the low-density outflow in
the polar region, as well as to the shock-heated remnant. Hence, most of
the matter is expected to have rather low values of the entropy, as shown
by the entropy distributions in the top panels of
Fig.~\ref{fig:ejecta-entropy}, which have a pronounced peak at very low
values, \ie $s \lesssim 10 \,{\kB}/{\rm baryon}$, and exhibit a sharp
decline for larger values. Note also that the tail to higher entropies
that is present in the quasi-circular models, is completely absent for
both EOS. This is again because most of the matter ejected in these
eccentric binaries is at low temperature.

As for Fig. \ref{fig:ejecta-ye}, also the top-right panel of
Fig.~\ref{fig:ejecta-entropy} reports the entropy distributions in the
case of unequal-mass binaries. Note that there does not appear to be any
systematic trend towards the distributions measured with quasi-circular
binaries, which is suggestive of the fact that shock heating in the
ejecta is strongly suppressed.

As mentioned above, Table~\ref{tab:model-results} reports, among other
quantities, also the average (median) values of the electron fraction and
of the entropy $\langle Y_{e} \rangle$ and $\langle s \rangle$. An
investigation of the results in the table reveals that the dynamical
ejecta are dominated by cold and neutron-rich matter, which is ejected
during the close passages at periastron by tidal torques. Any further
ejection by shock-heating is negligible, as long as one excludes those
mergers that occur with very small impact parameter and hence represent
essential ``head-on'' collisions. These conclusions are robust against
variations in the EOS, the periastron parametrization and the mass ratio
of the system.

\begin{figure*}
  \begin{center}
    \includegraphics[width=1.8\columnwidth]{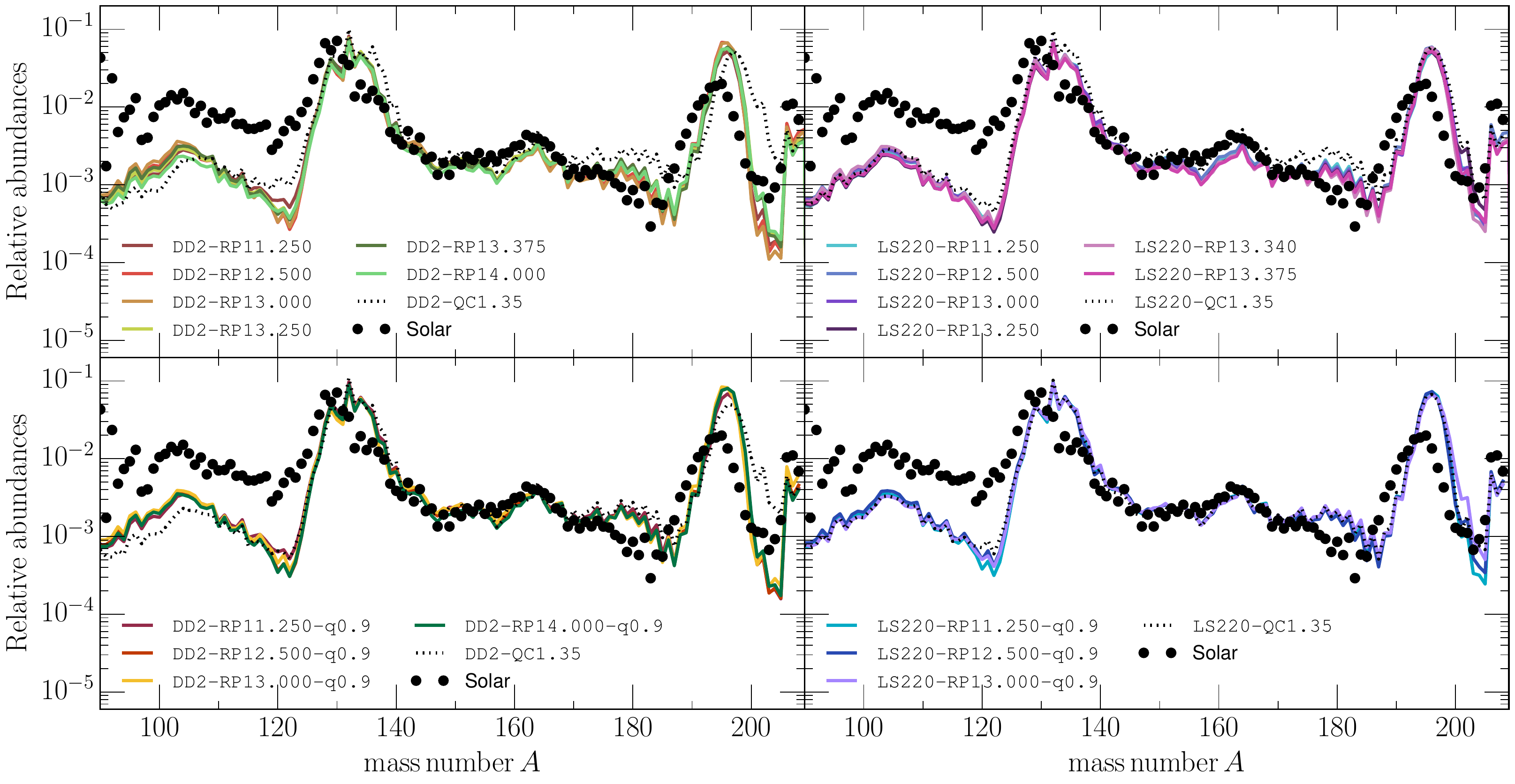}
    \caption{Final abundances of the r-process yields from a representative
      subset of unbound tracer trajectories for the equal-mass and $q = 0.9$
      binaries listed in Table~\ref{tab:model-summary} together with the
      solar abundances shown as black circles. We also show the abundances
      resulting from the tracers of the quasi-circular mergers of the same
      mass and EOS from \cite{Bovard2017}.}
    \label{fig:yields}
  \end{center}
\end{figure*}

Finally, we show in the bottom panels of Fig.~\ref{fig:ejecta-entropy}
the distributions of the velocities of the ejecta, which is the third key
property of the ejected material and plays an especially important role
in calculations of lightcurves from kilonova emission.  Note that in a
way similar to the distributions of quasi-circular binaries, the ejecta
exhibit velocities of at least $0.15$ and strongly peaks after that,
leading to average velocities of $0.21 - 0.25$ and a long tail of
velocities of up to $0.8$.  Note also that because smaller impact
parameters lead to larger shocks and hence to larger accelerations, the
distributions show an increase in the slow component and a decrease in
the fast component of the ejecta as the impact parameter is
increased. Indeed, the loss of a high-velocity tail is particularly
evident in the case of the DD2 binaries. Finally, this overall behaviour
does not change considerably when considering unequal-mass binaries, as
shown in the bottom-right panel of Fig. \ref{fig:ejecta-entropy}.  In
Table~\ref{tab:model-results} we report the mass-averaged ejecta velocity
$\langle v_{\rm ej}\rangle$, and its asymptotic value, $\langle
v_{\infty}\rangle$, \ie the value it would assume considering the
reduction of the kinetic energy to energy from the gravitational field of
the system. Such a velocity is computed assuming a point-mass field
configuration with ADM mass $M_{\rm ADM} = M_{1} + M_{2}$, where
$M_{1,2}$ are given in Table~\ref{tab:model-summary} [\cf Eq. (9) of
  Ref. \cite{Bovard2017}].

\subsection{r-process nucleosynthesis}
\label{sec:nucleo}

The study of the ejected matter is particularly important because of its
nuclear evolution once it has left the system and undergoes
nucleosynthetic reactions via the r-process. In our approach, the
r-process yields are computed using the setup of tracer trajectories
described in Ref. \ref{sub:tracers_nuc} and in which a subset of the
complete set of tracers is selected following the prescription described
in \cite{Bovard2017} to avoid the need of running the nuclear network for
every single tracer trajectory and to define an unambiguous mass weight
for each of them. In practice, our use of tracers and their input in the
nucleosynthetic analysis can be summarised as follows. First, the unbound
tracers are collected from all tracers of a simulation based on the
geodesic criterion. Second, instead of selecting a subset randomly from
the unbound tracers, we randomly select a representative tracer for a
specific value of the electron fraction and of the specific entropy. To
accomplish this we take the $Y_{e}$ and specific-entropy distributions of
the ejecta at the $200\,\Msol$ detector and bin them into a
two-dimensional histogram. Using then the knowledge of these two
quantities for every tracer at the moment it passes the detector, it is
possible to sample one tracer for each bin. The mass weight of each
tracer is naturally the mass fraction of the bin it was sampled
from. This prescription ensures that the whole range of possible
combinations are properly covered and weighted accordingly. Using this
procedure, each model effectively leads to $\sim 2000$ tracer
trajectories that are then trimmed and extended as described in
Sec.~\ref{sub:tracers_nuc}.

Furthermore, since we wish to carry out a comparison with the
nucleosynthetic yields and r-process abundances from quasi-circular
orbits computed in Ref. \cite{Bovard2017}, and since the latter employed
a different nuclear-reaction network code, we apply the same procedure
described above also to tracers computed from quasi-circular binaries. In
other words, instead of comparing the abundance curves computed here with
those reported in \cite{Bovard2017}, the r-process abundances from
quasi-circular binaries are recomputed with the same setup employed in
this paper. In this manner, we eliminate any possible bias effect related
to differences in the nuclear rates or nuclear-reaction network codes and
can thus perform a meaningful comparison.

\begin{figure*}
\begin{center}
\includegraphics[width=1.8\columnwidth]{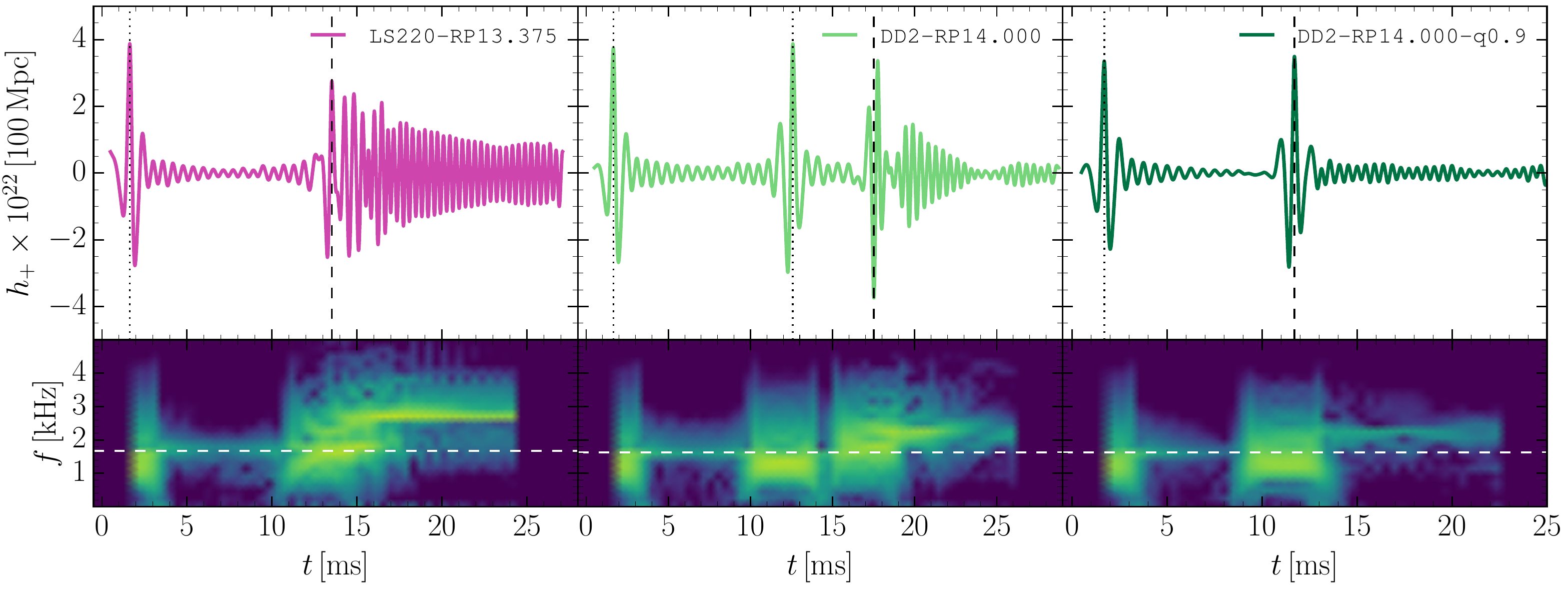}
\caption{Gravitational-wave strain of the $\ell = m = 2$ mode for a
  source at $\rm 100 \,{\rm Mpc}$ distance and its spectrograms for
  different $r_{p}$, EOS and $q$. Shown with a white dashed line are the
  respective $f$-mode frequencies while the time of periastron approach
  is marked with a vertical dotted and the time of merger with a vertical
  dashed black line. Note that there is significant power from the
  $f$-mode after the pericenter burst, which is well separated in time,
  from the contributions at similar frequencies during merger.}
\label{fig:gws-combined}
\end{center}
\end{figure*}

\begin{figure*}
\begin{center}
\includegraphics[width=1.8\columnwidth]{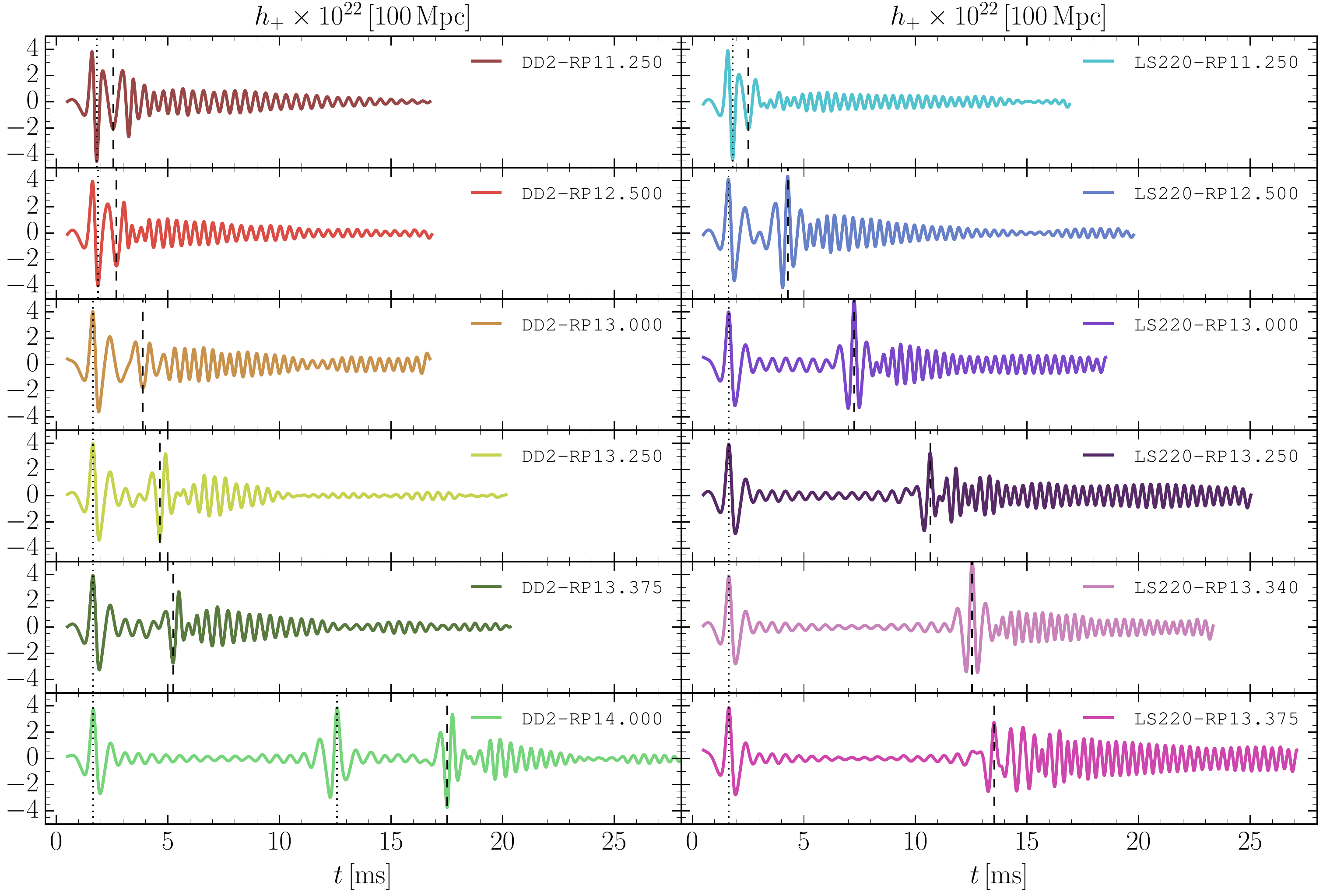}
\caption{Gravitational-wave strain for a source at $\rm 100 \,{\rm Mpc}$
  distance. The time of periastron approach is marked with a vertical
  dotted and the time of merger with a vertical dashed black line.}
\label{fig:gws-strain}
\end{center}
\end{figure*}

\vskip 2.0cm

The final abundance patterns of the r-process of all models considered in
this work and as computed with the nuclear-reaction network code
\texttt{SkyNet} are shown in Fig.~\ref{fig:yields} as a function of the
atomic mass number $A$. These abundances are supplemented by the
recomputed quasi-circular (dotted black lines) and solar r-process
abundances (black solid circle). All curves are normalized to unity in
the given mass number interval to eliminate the need of choosing a
special but arbitrary element to normalise to. The top panels refer to
binaries with equal masses (one for each of the two EOSs considered),
while the bottom panels show the results for unequal-mass binaries, again
for the two EOSs. In all cases, the different lines refer to different
values of the initial periastron parametrization.

Overall, the results reported in Fig. \ref{fig:yields} clearly show that
all models considered in this work are able to reproduce the general
behaviour of the solar r-process element abundances, starting from the
\textit{second} r-process peak at $A \simeq 130$ and up to the
\textit{third} peak at $A\simeq 195$, including the rare-earth elements
in between\footnote{Note that the low-$Y_e$ material ejected in our
  simulations is not able to reproduce the \textit{first} peak of the
  r-process elements. Indeed, in order to reproduce the whole range of
  r-process elements the ejection of matter at both low and high $Y_e$ is
  necessary, the latter being responsible for the abundances in the first
  peak of the distribution. Such high-$Y_e$ material could come, for
  instance, from the secular ejecta \cite{Wanajo2014}.}. Similar
robustness of the r-process against other astrophysical aspects in binary
neutron-star systems were found in earlier investigations, both numerical and
analytical, involving quasi-circular systems \cite{Lattimer74, Meyer89,
  Freiburghaus:1999, Korobkin2012}. Furthermore, while
differences in the properties of the ejecta from quasi-circular and
eccentric binaries have been discussed in Sec.~\ref{sub:dynejec-props},
these do not really impact the final abundances, which are found to
reproduce equally well Solar abundances for both classes of binaries. In
summary, these results confirm the findings of previous works and confirm
that the merger of binary systems of neutron stars can reproduce
robustly, \ie in a way that is insensitive to the details of the EOS,
mass ratio, and orbital dynamics, the Solar abundances of heavy r-process
elements.

\begin{figure*}
\begin{center}
\includegraphics[width=1.8\columnwidth]{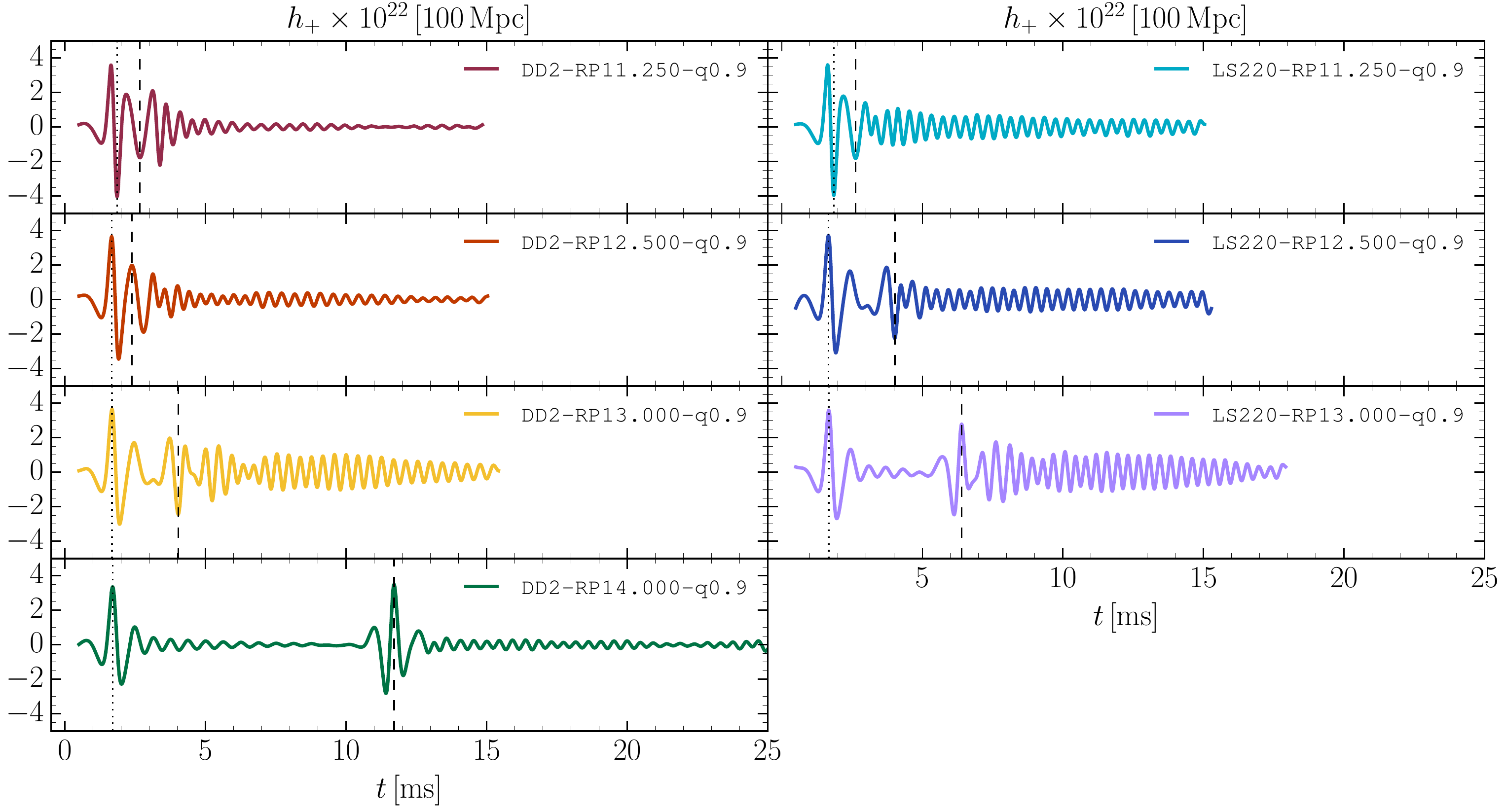}
\caption{The same as in Fig. \ref{fig:gws-strain} but for the
  unequal-mass binaries.}
\label{fig:gws-strain-q}
\end{center}
\end{figure*}
%

\subsection{Gravitational-wave emission}
\label{sec:ge}
%
The gravitational waveforms computed from our eccentric binaries and
reported\footnote{Note that the duration of the waveforms in
  Figs.~\ref{fig:gws-combined}--\ref{fig:gws-strain-q} is different for
  different binaries as the simulations for binaries with multiple
  encounters obviously require longer integration times.} in
Figs.~\ref{fig:gws-combined}--\ref{fig:gws-strain-q}, follow the same
overall behaviour observed in previous investigations of \cite{Gold2012,
  East2013}. More specifically, at close passages the orbital dynamics of
the binary gives rise to the GW bursts that are typical of highly
eccentric binaries. Such bursts are then separated in time by the GWs
associated with those oscillations that have been excited at the
close-periastron passage and that reflect the response of the stars to
the fundamental mode of oscillation, or $f$-mode. This is shown in
particular in Fig.~\ref{fig:gws-combined}, which reports in the top
panels the gravitational waveforms for the set of three binaries at
a distance of $100\,{\rm Mpc}$ with particular long waveforms, while
it offers a view of the corresponding spectrogram
\footnote{To produce the spectrogram we apply a Tukey window with a shape
  parameter of $0.25$, a window length of $4.5 \, \rm ms$ and an overlap
  of $4.0 \, \rm ms$.}  in the lower panels. Indicated with a horizontal
dashed white line is the frequency of the $f$-mode oscillation of each
individual neutron star model in isolation that is triggered after the
first encounter (before merger) while the main $f_2$ peak frequency of
the HMNS during the post-merger signal can be easily deduced from the
spectrogram (see \cite{Takami2014, Takami2015, Rezzolla2016} for a
classification of the spectral features of the post-merger signal). The
spectrograms reveal more clearly what can already be seen from the
waveforms, namely, that a portion of the energy in the signal of the
order of $\gtrsim 5\%$ lies in narrow frequency bands corresponding to
the $f$-modes of oscillation of the individual stars, which are then
followed by the oscillations of the HMNSs after the merger; similar
conclusions have also been obtained very recently by
\cite{Chaurasia2018}. Since these $f$-mode oscillations depend on the
internal structure of the stars, their measurement would yield an
independent avenue to constrain the neutron-star mass in addition to the
information from the inspiral and/or post-merger waveform thereby
breaking parameter degeneracies and allowing a more accurate
determination of source parameters if accurate templates are available.

Since their damping is mostly due to the GW emission, these oscillations
can be long-lived, so that at the onset of merger the waveforms may still
exhibit features due to eccentricity depending on details of the orbit
\cite{Gold2012,Gold:2012tk}. When considering the various binaries
reported in Fig. \ref{fig:gws-strain} (which refers to equal-mass
binaries) and focusing on the periastron approach (which is marked with a
vertical dotted black line) and to the time of merger (which is marked
with a vertical dashed black line), it is easy to appreciate that as the
impact parameter is increased from top to bottom, the binaries experience
either a single periastron approach (top panels), two periastron
approaches (middle panels), or even three periastron approaches
(bottom-left panel).

Note also that when keeping fixed the EOS and the impact parameter $r_p$,
the strongest tidal interactions occur in the equal-mass systems. This
can be most easily deduced when comparing the middle and right panels of
Fig. \ref{fig:gws-combined}, which report the DD2 binaries with
$r_p=14\,\Msol$ and mass ratios $q=1$ and $q=0.9$, respectively. In
particular, it is easy to see that the $f$-mode oscillations triggered
after the first encounter are larger and longer-lived in the equal-mass
case. This is because although the tidal torques are not necessarily the
largest\footnote{For extended objects such as neutron stars, the torque
  will depend sensitively on the mass, the mass-radius relation and the
  orbital frequency, all of which depend themselves on the EOS. It is
  therefore possible that the maximum torque is not attained in the case
  of equal-mass binaries.}, in the equal-mass case the two excited stars
resonate at exactly the same frequency and hence produce a stronger GW
signal (middle panel). By contrast, in the unequal-mass case (right
panel), the two stars oscillate at slightly different frequencies and the
corresponding GW signal is rapidly suppressed. More specifically, since
the two stars are physically separated and gravitational waves couple to
matter only weakly, the coupling between the oscillations in the two
stars is negligible in practice. At the same time, the gravitational-wave
emission relative to the $f$-mode and reported in the figure is the
result of the superposition of the distinct emissions from the two stars
and this is clearly larger when the two stars emit in phase.

\begin{figure*}
\begin{center}
\includegraphics[width=0.7\textwidth]{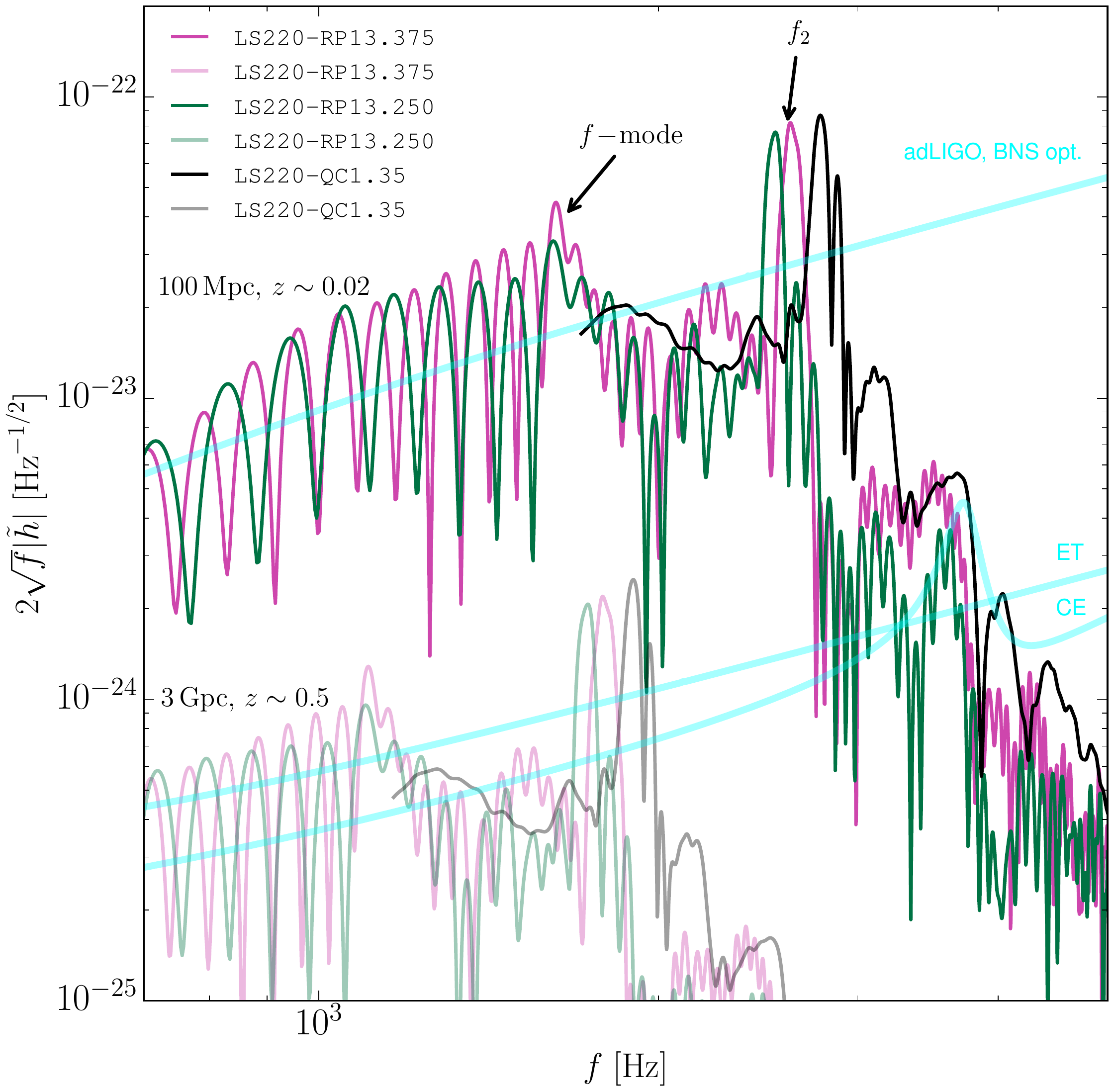}
\caption{PSDs from some of our binaries at a distance of $\sim 100\,{\rm
    Mpc}$, \ie $z=0.023$, reported with the same colorcode used in
  Figs. \ref{fig:gws-combined}--\ref{fig:gws-strain-q}, together with the
  detector sensitives for adLIGO, ET and CE at design sensitivity. Also
  shown as a comparison is the corresponding signal for a binary in
  quasi-circular orbit and with the LS220 EOS. Note that a good portion
  of the power is located in a somewhat broad bump around $1\,{\rm kHz}$
  with a high-frequency modulation due to the short duration of the
  signals and to the bursts at the periastron and then at the
  merger. Note also the peaks associated with the $f$-mode oscillations,
  as well as the $f_2$ peaks of the post-merger signal. Finally, reported
  with the same colours but semitransparent lines, are the same PSDs when
  the sources are taken to be at a larger distance of $3\,{\rm Gpc}$, \ie
  $z=0.5$.}
\label{fig:strain-detectors}
\end{center}
\end{figure*}

For the models shown in
Figs.~\ref{fig:gws-combined}--\ref{fig:gws-strain-q} -- whose first
periastron encounter is again marked with a vertical dotted black line,
while merger with a dashed black line -- the post-merger waveforms
exhibit the well-known features associated with an HMNS, with spectral
features that show clear peaks \cite{Bauswein2011, Stergioulas2011b,
  Takami2014, Takami2015, Bernuzzi2015a, Palenzuela2015}, following
universal relations in the case of binaries in quasi-circular orbits
\cite{Takami2015, Rezzolla2016}\footnote{Note that these universal
  relations for the $f_2$ frequencies are not applicable in the case of
  eccentric binaries since the position of the $f_2$ peak now varies with
  the eccentricity of the orbit as a result of the different angular
  momentum of the merger remnant (\cf
  Fig. \ref{fig:strain-detectors}).}. Furthermore, it is clear from
Figs.~\ref{fig:gws-combined}--\ref{fig:gws-strain-q} that the LS220 EOS,
being slightly stiffer than the DD2 EOS, typically yields -- all else
being equal -- a stronger GW emission during the HMNS phase (see also
Ref. \cite{Rezzolla2016} for a detailed discussion). At the same time, it
should be noted that given the typical masses and radii of neutron stars,
and the fact that the fundamental mode frequency scales like the average
rest-mass density, \ie $f\simeq (M/R^3)^{1/2} \gtrsim 1\,{\rm kHz}$,
these frequencies generically fall in a range where the advanced GW
detectors [adLIGO and advanced Virgo, (adVIRGO)] have lower sensitivity
as they are limited by the laser shot-noise. Hence, future
third-generation detectors, such as ET or CE, or any detector with
recently suggested upgrades \cite{Miao2015,Miao2017}, are certainly
better suited for the detection of the GW signal from eccentric binaries.

The first detections of such rather rare systems are expected to come
from sources at larger cosmological distances
\cite{Messenger2013}. For quasi-circular binaries, the properties of
the GW emission lead to huge detector volumes out to $z \leq 3$ (ET)
and $z \leq 6$ (CE), also because the GW signal is shifted to
frequencies where the detectors are more sensitive. Detection efforts
of gravitational waves from eccentric binaries will also benefit from
the cosmological redshift of the frequency of either the $f$-mode
oscillations or that of the post-merger signal with intrinsic
frequencies of a few kHz in the source frame. Due mostly to the lack
of substantial power at lower frequencies, the sources studied here
can be seen out to more modest, but still large distances of about $z \gtrsim 0.5$.

\begin{figure*}
\begin{center}
\includegraphics[width=1.7\columnwidth]{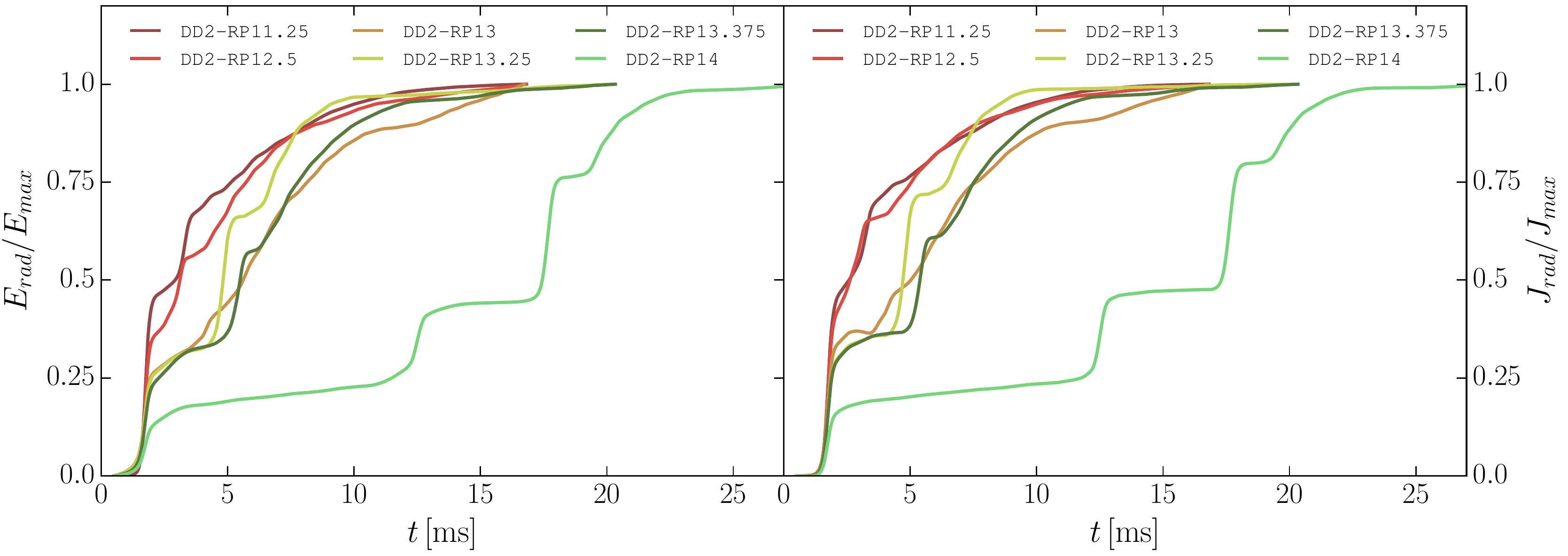}
\caption{Energy $E_{\rm rad}$ and angular momentum $J_{\rm rad}$ with
  respect to the total energy $E_{max}$ and total angular
  momentum $J_{max}$ radiated by GWs as a function of time for
  different values of $r_{p}$ for the DD2 EOS.}
\label{fig:gws-energy}
\end{center}
\end{figure*}

To illustrate this effect, we compare in Fig.~\ref{fig:strain-detectors}
the PSDs from some of our models to the noise equivalent strain due to
the detector noise for adLIGO, ET and CE at design sensitivity
considering a source distance of $\sim 100\,{\rm Mpc})$, \ie $z=0.023$,
marking the different binaries with the same colorcode used in the
previous figures \ref{fig:gws-combined}--\ref{fig:gws-strain-q}. Also
shown as a comparison is the corresponding signal for a binary in
quasi-circular orbit and with the LS220 EOS. Note that for the eccentric
cases a good portion of the power is located in a somewhat broad bump
around $1\,{\rm kHz}$ with a high-frequency modulation that is most
likely due to the short duration of the signals and to the bursts at the
periastron and then at the merger (see Appendix \ref{sec:appendix} for a
discussion with a toy model). Also clearly shown in the PSDs are the
peaks associated with the $f$-mode oscillations, as well as the $f_2$
peaks of the post-merger signal\footnote{The emission from the HMNS
  oscillations was followed only for the timescale of the simulations but
  could actually be much more long-lived, increasing the signal power at
  least as the square root of the emission time.}.

When comparing the PSDs from eccentric and quasi-circular binaries it is
apparent that the latter benefit from a considerable amount of power
accumulated during the inspiral, which is much reduced in the case of
eccentric mergers and, more importantly, does not follow a simple
$f^{-7/6}$ behaviour with frequency and that makes it amenable to
searches with GW templates. Indeed, bursts-type searches based on
unmodeled features and an extensive use of spectrograms such as those
shown in Fig. \ref{fig:gws-combined}, are possibly the most effective way
of searching for this type of signals. Overall, the lack of this
low-frequency portion reduces the overall SNR and implies that
dynamical-capture signals will be in general harder to detect than
quasi-circular binaries with the assumed detector characteristics (all
else being equal).

Also shown in Fig.~\ref{fig:strain-detectors} with the same colours but
semitransparent lines, are the same PSDs when the sources are taken to be
at a larger luminosity distance of about $3\,{\rm Gpc}$, corresponding to
redshift of $z=0.5$. The gravitational redshift of which the signal will be
subject to in this case, and the corresponding decrease in power due to
the larger distances, will obviously make it undetectable by adLIGO, but
still within the detection horizon of third-generation detectors such as
ET or CE, despite the smaller power at low-frequencies contributing to
the SNR.

In summary, despite the difficulties of detecting the GW signal from such
systems, the combined benefits of having a very large detection volume
(which counters the low event rate) and the shift to lower frequencies of
the bulk of the emission (which counters the low sensitivity of the
detectors), considerably increases the chances of detecting the GW
emission from eccentric binaries by third-generation detectors.

Finally, we show in Fig. \ref{fig:gws-energy} the energy, $E_{\rm rad}$,
and angular momentum, $J_{\rm rad}$, radiated by GWs as a function of
time for different values of the initial impact parameter and for the
equal-mass binaries with the DD2 EOS. As can be appreciated from the two
panels, the efficiency in the emission of GWs is very large around the
time of closest approach, when the actual luminosity reaches values of
$\simeq 5 \times 10^{55} \, {\rm erg/s}$, and where large fractions (up
to $50\%$) of the total radiated energy and angular momentum are
emitted. Note also that the GW waveforms associated with these bursts can
be easily modeled by sine-Gaussian and hence their presence in the
detector's signal can easily and efficiently be matched-filtered to be
revealed. Furthermore, it is important to note that for a given EOS the
total radiated energy does not have a simple monotonic behaviour with the
initial impact parameter. More specifically, while the brightness of the
GW burst decreases with $r_p$ for binaries having only one encounter --
so that, for instance, the binary \ttt{DD2-RP11.250} radiates more than
\ttt{DD2-RP13.375} -- the total radiated energy increases with the number
of encounters, so that at the end the binary \ttt{DD2-RP14.000} is
actually more luminous than \ttt{DD2-RP11.250}. Finally, note that the
time intervals between two encounters, when the $f$-mode oscillations are
excited, the GW emission is steady and the radiated energy and angular
momentum are simple and almost linear functions of time.

\section{Conclusions}
\label{sec:conclusions}

We have presented the most extensive investigation to date of eccentric
binary neutron-star mergers in full general relativity and encompassing
19 different configurations. In this way, we were able to explore the
impact of various degrees of freedom of the system -- the orbital
eccentricity (or impact parameter), the EOS and the mass ratio -- on
the bi-products of the merger, namely, the GW emission, the dynamically
ejected matter, and the nucleosynthetic yields.

Our numerical approach is based on the CCZ4 formulation of the Einstein
equations, whose constraint-damping capabilities make it particularly
suited to study a system involving initial data with considerable
violation of the constraints. For the dynamically ejected mass, on the
other hand, we employ a standard criterion for unbound material and track
it via tracer particles that are passively advected with the fluid
flow. We use a novel approach to seeding the tracer particles both at the
first periastron and first apoastron respectively to better capture the
peculiarities of eccentric binaries, while minimising potential
tracer-selection bias. These resulting diagnostic quantities from these
tracers are then interfaced with the separate \texttt{SkyNet}
nuclear-reaction network code that produces nucleosynthetic yields.

Overall, our results indicate that eccentric binary neutron-star mergers
universally produce considerably more ejected mass, \ie
$10^{-2}-10^{-1}\,\Msol$. Furthermore, these ejecta are faster (with long
tails of the distribution up to $0.8$) and with a composition that
depends on the EOS, but is distinct from that of merging systems on
quasi-circular orbits. Also quite generically, we can identify a
component of the dynamically ejected material that is lost near the
equatorial plane and that is characterised by much lower temperatures and
low $Y_e$ as a result of the very small shock heating experienced by this
matter. On the other hand, matter ejected around the polar regions does
not appear to be a promising site for the nucleosynthesis of the heaviest
r-process elements due to its relatively high values of $Y_e$, high
temperatures and low densities.

Despite some of these striking differences between eccentric and
quasi-circular binaries, the final nucleosynthetic yields do not
depend on the EOS and reproduce the main features of the observed
relative abundances and, in particular, the second and third peak of the
heavy-elements distribution. This is a particularly important result:
since a similarly good match with the observations has been obtained by
several authors when considering quasi-circular binaries, it is quite
apparent that the merger of binary neutrons stars leads to a
nucleosynthesis that reproduces the observed abundances with little
dependence on the orbital dynamics, on the EOS, the total mass and the
mass ratio.

Finally, we have computed the GW emission focusing on $f$-mode
oscillations before and after merger finding that the imprints of
$f$-mode oscillations in the GW signals depend most strongly on the
orbital configuration and mass ratio of the binary, but are rather
uniform across different EOSs. We attribute differences to the appearance
of $f$-mode features in GWs to the different tidal interactions both in
terms of excitation and response. In our analysis we also include the
cosmological redshift of the GW signal from large distances, and illustrate
how this brings the signal into a more sensitive regime of
third-generation detectors. We also show that, while some of the systems
under investigation can be very challenging to detect even for the ET or
CE, other configurations with longer signals, more cycles and
low-frequency contributions are detectable out to distances as large as
$z \gtrsim 0.5$, despite the lack of low-frequency cycles that are such an
important contribution in quasi-circular signals. Detector sensitivity at
high frequencies around $1\,{\rm kHz}$ are key to detecting such
sources.

Various effects have not been taken into account in this work, such as
magnetic- and neutrino-driven winds, viscous heating in the surrounding
tori and magnetic fields, which operate on longer timescales can increase
the ejected mass \cite{Dessart2009, Fernandez2013, Fernandez2015,
  Siegel2014, Metzger2014, Rezzolla2014b, Ciolfi2014, Just2015,
  Martin2015, Kiuchi2015a}. In particular, it has been shown in
Ref. \cite{Radice2016} that neglecting neutrino cooling can lead to an
overestimation of the ejected mass. Another important process could be
neutrino heating modeled by an M1-scheme \cite{Sekiguchi2015,
  Sekiguchi2016}. This could lead to even more unbound material as has
been shown in \cite{Foucart2015}. All of these aspects of the dynamics of
merging eccentric binaries will find improvement in future work.

\section*{Acknowledgements}

It is a pleasure to thank Luke Bovard, Jonas Lippuner, Moritz
Reichert, Marius Eichler, Elias Most, Charles Horowitz, Huan Yang, and
Eric Poisson for numerous discussions and useful input. Support comes
also in part from ``NewCompStar'' and ``PHAROS'' , COST Actions MP1304
and CA16214; LOEWE-Program in HIC for FAIR; European Union's Horizon
2020 Research and Innovation Programme (Grant 671698) (call
FETHPC-1-2014, project ExaHyPE), the ERC Synergy Grant ``BlackHoleCam:
Imaging the Event Horizon of Black Holes'' (Grant No. 610058). The
simulations were performed on the SuperMUC cluster at the LRZ in
Garching, on the LOEWE cluster in CSC in Frankfurt, on the HazelHen
cluster at the HLRS in Stuttgart.

\appendix

\section{On the appearance of multiple bursts containing similar frequencies in GW spectra}
\label{sec:appendix}

In this Appendix we illustrate through several simple analytical toy
models that mimic the basic features in our waveforms, how GWs with more
than one burst can lead to non-continuous PSDs. We build the toy
model combining various contributions from the following elements, that
are representative of features we observe in our gravitational
waveforms. In particular, given a kernel
\begin{align}
b(t, t_{burst},\sigma,\omega) :=& \exp{\left[-(t-t_{burst})^2/\sigma \right]}
\sin(\omega t)\,,
\end{align}
the signals are then explicitly formed as
\begin{align}
s_1 :=&\,  b(t,10^3,10^5,1/20) + (1/10) \sin(t/100) + \label{eq:2bursts} \\
 &\,   + b(t,4\times10^3,8\times 10^4,1/22)\,, \nonumber \\
s_2 :=&\,  b(t,10^3,10^5,1/20) + (1/10) \sin(t/50)\,, \label{eq:1burst+sin} \\
s_3 :=&\,  e^{-(t-10^3)^2/5\times 10^4} + e^{-(t-3\times10^3)^2/5\times 10^4}\,. \label{eq:2gaussians}
\end{align}
where $t$ is uniformly spaced on the interval $[0,5000]$ using $10^4$
points.

\begin{figure}[thb]
\begin{center}
\includegraphics[width=1.0\columnwidth]{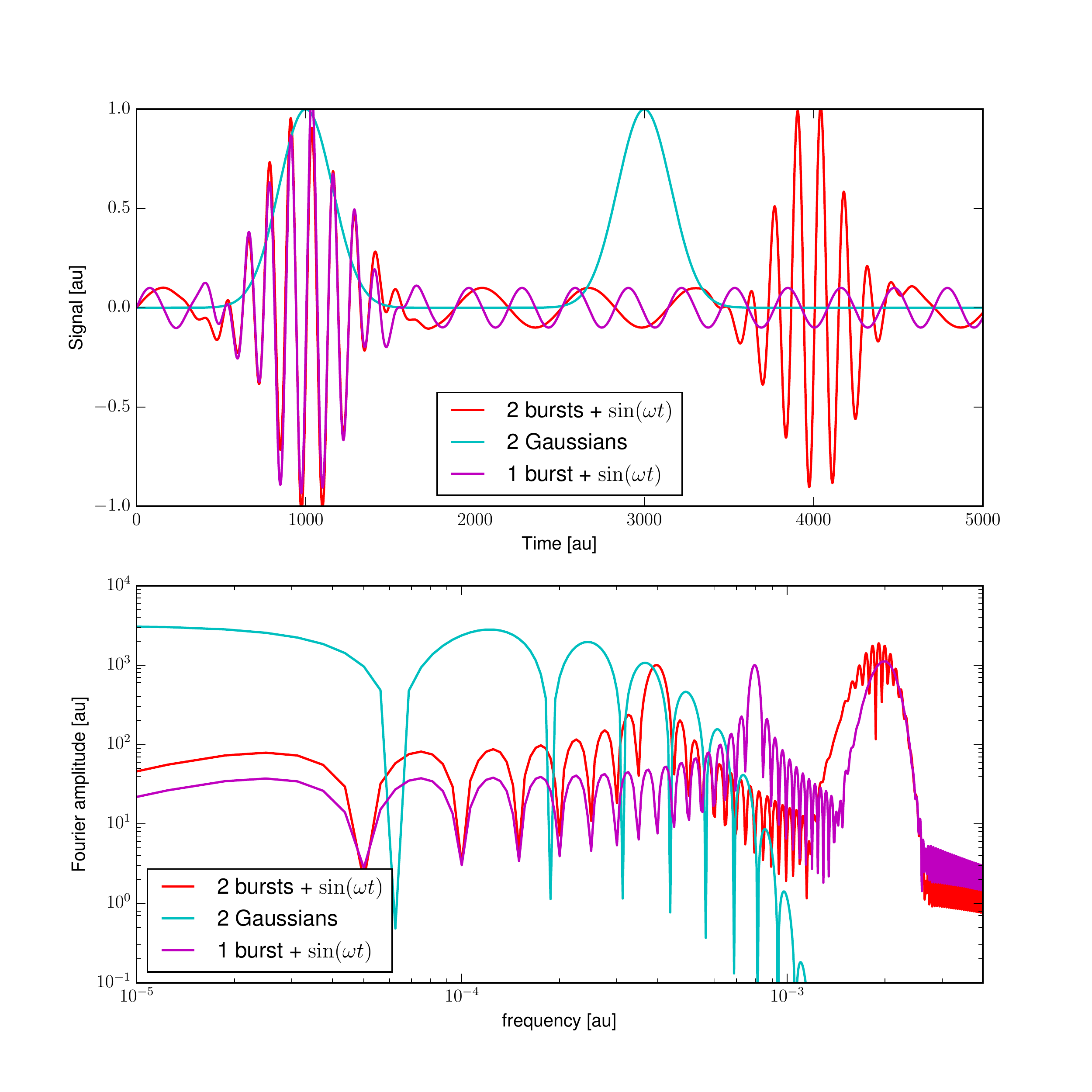}
\caption{Three toy-model signals illustrating well-known relations
  between the time domain and frequency domain: A sinusoidal signal with
  one (magenta) or two (red) bursts and a signal composed of two pure
  Gaussians with no periodic component (cyan). All signals are
  zero-padded such that there length increases by a factor of $8$.}
\label{fig:zigzag-toys}
\end{center}
\end{figure}

Equation~\eqref{eq:2bursts} constitutes a model with two bursts, as well
as a sinusoidal wave analogous to the $f$-mode signal. This signal,
termed "$2\ {\rm bursts} + \sin(\omega t)$" in
Fig.~\ref{fig:zigzag-toys}, exhibits a non-continuous structure where
regions of high power are separated by small (in frequency) regions with
very low power.  This behaviour is very similar to our eccentric GW PSDs
in Fig.~\ref{fig:gws-combined}, despite the fact that our toy models are
fully analytic and smooth functions of time. This implies that the
non-continuous structure shown in Fig.~\ref{fig:gws-combined} is not an
artefact due to the poor sampling of the GWs. Interestingly, also other
toy models show the same behavior as long as there are two bursts that
can interfere with each other. In contrast, models with only one burst
(even with an additional periodic component) are more akin to
quasi-circular waveforms do not show such non-continuous PSDs.

To further confirm this interpretation we have computed GW spectra from a
subset of the GW signals and consistently found that whenever we filter
out all bursts except for one (either the one at periastron or at merger)
the non-continuous structure disappears, as expected based on the
analysis summarised above.

\bibliographystyle{apsrev4-1}
%

\end{document}